\documentclass[aps,prd,12pt,nofootinbib,floatfix]{revtex4}
\usepackage{graphicx}
\usepackage[font=small,skip=0pt,justification=raggedright]{caption}
\usepackage{amsmath}
\usepackage{amssymb}
\usepackage{mathrsfs}
\usepackage{verbatim}
\usepackage{dutchcal}

\usepackage[normalem]{ulem}
\usepackage{xcolor}

\newcounter{fig}   \newcommand{\lbfig}[1]{\refstepcounter{fig}
\label{#1} }

\newcommand{\bea}{\begin{eqnarray}}
\newcommand{\eea}{\end{eqnarray}}
\newcommand{\be}{\begin{equation}}
\newcommand{\ee}{\end{equation}}

\newcommand{\re}[1]{(\ref{#1})}

\newcommand{\eqn}{\begin{eqnarray}}
\newcommand{\eqnx}{\end{eqnarray}}

\begin{document}

\title{\boldmath Stress stability criterion of $U(1)$ gauged non-topological solitons in the 3+1 dimensional O(3) sigma-model \unboldmath}

\author{Aliaksei Mikhaliuk}
\email{alekseybsu.mihalyuk@gmail.com}
\affiliation{Belarusian State University, Minsk 220004, Belarus}

\author{Yakov Shnir}
\email{shnir@theor.jinr.ru}
\affiliation{BLTP, JINR, Dubna 141980, Moscow Region, Russia}

\date{\today}

\begin{abstract}
We study the energy-momentum tensor of the spherically symmetric non-topological solitons of the  $O(3)$ non-linear sigma-model with a standard kinetic term and with a symmetry breaking potential in 3+1 dimensional flat space-time. We evaluate the distributions of the corresponding energy density, shear forces and pressure and study the stability criteria for these solutions.
We argue that the presence of domains with negative energy density and violation of the energy conditions most likely do not lead to destabilization of solitons. 
\end{abstract}

\maketitle

\newpage

\section{Introduction}

Much attention has been devoted over the last few decades to the study of soliton solutions which arise in various non-linear systems. They represent regular localized lumps of energy. These  field configurations emerge 
in many different areas of physics, e.g., physics of condensed
matter 
\cite{Solitons,Ackerman:2017lse}, solid state physics \cite{Kosevich}, non-linear optics \cite{Mollenauer}, biophysics \cite{Dauxois} and cosmology \cite{Vilenkin}. 
In many cases existence of
the solitons is related with topological properties of the
system; see, e.g., \cite{Manton:2004tk,Shnir2018}.
There are two different classes of soliton configurations: topological and non-topological solitons. While topological solitons are characterized by a certain topological invariant related to the vacuum structure of the theory, non-topological solitons possess a conserved Noether charge related to some symmetry of the Lagrangian of the model. However, there are theories that support both types of solitons.

An interesting example of such a model is the well-known nonlinear sigma model \cite{Gell-Mann:1960mvl} and its modifications. The field of the $O(3)$ model is a triplet $\phi^a=(\phi^1,\, \phi^2,\, \phi^3)$ restricted to a two-sphere, $(\phi^a)^2=1$, it asymptotes to the vacuum 
$\phi_\infty^a=(0,0,1)$. In the simplest 2+1 dimensional case, corresponding soliton solutions belong to an equivalence class characterized by the homotopy group    $\pi_2(S^2)=\mathbb{Z}$. Soliton solutions of this conformally invariant self-dual planar theory without a potential were constructed in \cite{Polyakov:1975yp}.

Further, in 3 spacial dimensions there also are topological soliton solutions of the $O(3)$ model   
classified by the linking number, the first Hopf map
$\phi^a: S^3 \mapsto S^2$. However, according to the 
Derrick’s theorem \cite{Derrick:1964ww},
to support stable soliton configurations in 3+1 dimensions, the Lagrangian of the $O(3)$ model should also
include, apart from the usual kinetic
term, a Skyrme-type term which is quartic in derivatives \cite{Faddeev:1976pg,Gladikowski:1996mb,Battye:1998pe}. In addition, a
potential term, which does not contain derivatives, 
can be included in the Lagrangian.  Although
the structure of the potential term of the Faddeev-Skyrme model is largely arbitrary, its particular choice determines different ways of symmetry breaking, see e.g.,  \cite{Harland:2013uk,Battye:2013xf,Samoilenka:2017hmn}.

Another mechanism of stabilization of solitons, which allows to circumvent Derrick's theorem,
is related with internal rotations of the configuration. Such time-dependent solitons were constructed
both in the planar $O(3)$ model with a symmetry breaking potential \cite{Abraham:1991ki,Ward:2003un}, and in the
$\mathbb{C}P^2$ model \cite{Amari:2024pnw,Antsipovich:2025liy}. Evidently, there
is a certain similarity between  these internally rotating topological solitons \cite{Radu:2008pp}
and Q-balls, which are time-dependent
lumps of a complex scalar field with a stationary oscillating phase \cite{Rosen,Friedberg:1976me,Coleman:1985ki,Lee:1991ax}.

Note that,  in general, internally rotating stationary solitons cannot be approximated as a rigid body, moreover,  
topological arguments alone cannot 
guarantee stability of solitons,
they can shrink or expand unless balanced by other physical mechanisms and interactions.
The deformations of topological solitons may result in instability of the configuration with respect to decay into constituents \cite{Halavanau:2013vsa,Battye:2013tka}.
Depending of the structure of the potential and the
values of the parameters of the model, there can be stable, metastable and unstable solutions.

It was pointed out that static non-topological solitons may exist in 3+1 dimensions in the $O(3)$ sigma model with a scalar potential which includes a difference of two positive‐definite terms \cite{Verbin:2007fa}. Since such potentials are not being positive everywhere, one expects violation of the energy conditions in the system. A prototype of such potential is the well-known Lennard-Jones potential, which combines long-range and short-range interactions.  

Further modification of the potential allows for the existence internally rotating non-topological $O(3)$ solitons \cite{Ferreira:2025xey}, it makes such configurations similar to the usual Q-balls. 
Such potential breaks the $O(3)$ symmetry of the model to the $SO(2)$ subgroup, 
further,  by analogy with the similar
situation in the planar baby Skyrme model \cite{Bogolubskaya:1989ha,Bogolyubskaya:1989fz,Piette:1994jt,Piette:1994ug}
one can
gauge this Abelian subgroup and add the Maxwell term to the Lagrangian of the system \cite{Gladikowski:1995sc,Cavalcante:1996zg,Samoilenka:2018oil,Samoilenka:2017fwj,Samoilenka:2016wys,Samoilenka:2015bsf,Shnir:2014mfa,Ferreira:2025xey}. 
The presence of a long-range Coulomb field in the system can significantly affect the properties of the soliton, the configuration coupled to the Abelian gauge field shows critical thresholds based on coupling parameters.

The problem of stability of various solitons 
can be analyzed by studying of the matrix
elements of energy-momentum tensor and related spatial distributions of the
forces acting in the interior of the configuration
\cite{Polyakov:2002yz,Polyakov:2018zvc,Mai:2012yc,Mai:2012cx,Loginov:2020xoj,Loiko:2022noq,Panteleeva:2023aiz,Farakos:2025byy}.
This approach supplements the Vakhitov-Kolokolov criteria of stability of solitons  \cite{Vakhitov:1973lcn}, 
it originates from the study of 
form factors of the energy-momentum tensor of hadrons \cite{Polyakov:2002yz,Polyakov:2018zvc} and evaluation
of the so-called "D-term", the quantity which is related to the mechanical deformations of the configuration. Note that
this method can be applied for both topological \cite{Panteleeva:2023aiz,Farakos:2025byy}
and non-topological \cite{Mai:2012yc,Mai:2012cx,Loiko:2022noq} solitons. More generally, all finite energy regular localized configurations
satisfy certain criteria for the distribution of the shear force and the pressure. Further, it has been shown that the electrostatic repulsion may destabilize Q-balls \cite{Loiko:2022noq}.  

In this paper we extend this approach 
to the case of $U(1)$ gauged spherically symmetric non-topological solitons of the $O(3)$ sigma model with a symmetry breaking potential in flat space. We study the energy-momentum tensor of the system and discuss corresponding distributions of the
pressure and shear forces, acting in the interior of the soliton. 

This paper is organized as follows. In Sec. II, we introduce the model 
including the field equations,  conserved Noether current and the stress-energy tensor. Scaling type arguments
are then used to establish the existence of regular electrically charged solitonic solutions. Here, we describe the spherically-symmetric parametrization of
the fields and the boundary conditions imposed on the configuration. 
The internal mechanical properties of the $U(1)$ gauged non-topological solitons of the $O(3)$ sigma model are discussed in Section III, here, in particular we study the shear forces and the pressure in the interior of these localized configurations. Here we also discuss the stability
condition of the solitons which follow from the conservation of the energy-momentum tensor of the systems. 
Numerical results are presented in Sec. IV.
Conclusions and remarks are formulated in Section V.

\section{The model and field equations}
\subsection{The action and the field equations}
We consider the $U(1)$ gauged non-linear $O(3)$ sigma-model in a (3+1)-dimensional Minkowski space-time, defined by the Lagrangian
density 
\be
L_{m} = -\frac14 F_{\mu\nu}F^{\mu\nu} + \frac12 (D_\mu\phi^a)^2 - U(\phi) \, ,
\label{Gauged_O3}
\ee
where the real triplet of the scalar fields $\phi^a$, $a=1,2,3$, is constrained as $(\phi^a)^2=1$, and $U(\phi)$ is a symmetry breaking potential. We
introduced the usual Maxwell term with the field strength tensor defined as $F_{\mu\nu}=\partial_\mu A_\nu - \partial_\nu A_\mu $. The covariant derivative of the  field $\phi^a$, that minimally
couples the scalar field to the gauge potential, is
\be
D_\mu \left(\phi^1+i\,\phi^2\right) = \partial_\mu \left(\phi^1+i\,\phi^2\right)
-i\,e\,A_\mu\left(\phi^1+i\,\phi^2\right)\,, \qquad D_\mu
\phi^3 = \partial_\mu \phi^3, 
\label{covariant}
\ee
with $e$ being the gauge coupling. 

Following previous discussions in  \cite{Verbin:2007fa,Ferreira:2025xey,Kunz:2025tsy},  we will consider the  potential $U(|\phi|)$, which breaks $O(3)$ symmetry to the subgroup $SO(2)\sim U(1)$ and is taken as the difference of two positive 
definite terms (see Fig.~\ref{fig0u}),

\begin{equation}
    U(\phi^a)=\mu^2(1-\phi^3)-\beta(1-\phi^3)^4\, ,
\label{pot}
\end{equation}
where $\mu$ and $\beta$ are two real parameters. Clearly, rescaling of the model $\beta \to \beta/\mu^2 $ allows us to set $\mu=1$.  

Importantly, the potential \re{pot}, apart from the common weakly attractive "pion mass" term, also contains a term that yields an additional short-range interaction
\cite{Leese:1989gi,Salmi:2014hsa,Samoilenka:2015bsf,Gillard:2015eia}. 
Potentials of that type may support localized non-topological soliton solutions in 3+1-dimensional flat spacetime \cite{Verbin:2007fa,Ferreira:2025xey},
they exist for $\beta > \beta_c=\frac{1}{8}$, as the potential \re{pot} possesses a global minimum at the anti-vacuum $\phi^3 = -1$, where $U(\phi) < 0$. The vacuum  $\phi^3 = 1$ corresponds to the local minimum, as $\beta=\beta_c=1/8$, the vacuum becomes degenerated, and the field configuration becomes an infinitely thin spherical wall separating two vacua\footnote{Note there is a significant difference between the soliton solutions of the non-linear $O(3)$ sigma model and the usual Q-balls in a complex scalar field theory \cite{Rosen,Friedberg:1976me,Coleman:1985ki,Lee:1991ax}, for which the vacuum value of the field is $\phi^3 = 0$.}.  Stable soliton solutions of the model \re{Gauged_O3}
do not exist as the rescaled parameter  $\beta$ becomes smaller than 1/8.

The model \re{Gauged_O3} is invariant with respect to
the local Abelian gauge transformations
\be
(\phi^1 + i \phi^2) \to
e^{ie \zeta}(\phi^1 + i \phi^2), \quad A_\mu \to A_\mu + \partial_\mu \zeta
\label{rotate}
\ee
where $\zeta$ is a real function of the coordinates. The corresponding Noether current
can be written as follows:
\be
j^\mu=\phi^1 D^\mu\phi^2 - \phi^2 D^\mu\phi^1
\label{ncurrent}
\ee

Variation of the Lagrangian \re{Gauged_O3}  with respect to
the fields $A_\mu$ and $\phi_a$ leads to the dynamical equations:
\be
\partial_\mu  F^{\mu\nu}=e  j^\nu,
\quad D^\mu D_\mu \phi^a + \phi^a( D^\mu \phi^b \, D_\mu \phi^b)+ \frac{\delta U}{\delta \phi}\phi^a =0\,  ,
\label{eqfield}
\ee
where we take into account the dynamical constraint imposed on the $O(3)$ field.
The vacuum boundary
conditions are 
$\phi_{\infty}^a = (0,0,1),\, D_\mu\phi^a =0,\, F_{\mu\nu}=0$. In the static
gauge, where the fields have no explicit dependence on time, the
asymptotic boundary conditions on the gauge potential are
\be
A_0(\infty)=V,\quad A_i(\infty)=0
\ee
where $V$ is a real constant. 
Note that the asymptotic value
of the electric potential $A_0(\infty)$
can be adjusted via the residual $U(1)$ transformations \re{rotate}, choosing  $\zeta =-Vt$.
In such a stationary gauge $A_0(\infty)=0$ and two components of the scalar triplet acquire an explicit harmonic time  dependence with frequency $\omega=eV$. 

The scalar and electromagnetic components of the energy-momentum tensor are
\begin{align}
    &T_\phi^{\mu\nu}=D^\mu\phi^a D^\nu\phi^a-\eta^{\mu\nu}\left(\frac12D^\rho\phi^a D_\rho\phi^a + U\right),\\
    &T_{em}^{\mu\nu}=F^{\mu\rho}F_{\rho}^\nu-\frac14 \eta^{\mu\nu}F^{\rho\sigma}F_{\rho\sigma}\, ,
    \label{tmunu}
\end{align}
respectively. Here  
$\eta_{\mu\nu}=~{\rm diag}\, (-1,1,1,1)$ is the Minkowski metric.

\subsection{Scale transformations and virial identity}

Let us establish that the model \re{Gauged_O3} with two-component potential \re{pot} supports finite energy localised solutions.  
The total energy functional of the system can be written as
\be
E=\int d^3 x \, \left(T_\phi^{00} + T_{em}^{00}\right) = E_0 + E_2 + E_4 \, , 
\label{eng}
\ee
where 
\be
E_0= \int d^3x  U(\phi), \quad
E_2=\frac12\int d^3x  D_i\phi^a D_i\phi^a ,\quad
E_4=\frac12 \int d^3x (E_i^2 + B_i^2)\, ,
\label{Derek}
\ee
and $E_i=F_{0i}$ and $B_i=\frac12 \varepsilon_{ijk}F_{jk}$ are the electric and magnetic fields, respectively. For the  potential \re{pot} $E_0$, which represents a  difference
of two positive definite terms, $E_0 = E_0^{(1)}-E_0^{(2)}$, where both $E_0^{(1)}$ and $E_0^{(2)}$
are positive-definite. 

The critical points of the total energy functional of the $O(3)$ sigma model \re{eng}
must satisfy the arguments of Derrick's theorem. Note that the corresponding scale transformation $x_i \to x_i^\prime =\lambda^{-1}x_i$ does not affect the field 
$\phi^a \rightarrow \phi^a$ because of the sigma model constraint,
and the gauge potential transforms as $A_\mu(x_i) \rightarrow A_\mu^\prime(x_i^\prime) =\lambda\, A_\mu(x_i)$. Hence,
\be
E(\lambda) =\lambda^{-1}E_2 + \lambda E_4 +
\lambda^{-3}E_0^{(1)} - \lambda^{-3}E_0^{(2)} 
\ee
and $\partial^2_\lambda E(\lambda=1)=2[E_2+6(E_0^{(1)} - E_0^{(2)})] \ge 0$. The corresponding virial identity
follows from the condition $\partial_{\lambda} E\mid_{(\lambda=1)}=0$, it gives
\be
E_2+3 E_0^{(1)}=E_4 + E_0^{(2)} \, ,
\label{virial}
\ee
which secures the existence of soliton solutions even in the absence of an electromagnetic field
\cite{Ferreira:2025xey,Kunz:2025tsy}. It should be noted that electromagnetic interaction in three spatial dimensions alone cannot stabilize a soliton. On the other hand, there are ungauged non-topological solitons in a model with a potential like \re{pot}  \cite{Verbin:2007fa,Ferreira:2025xey,Kunz:2025tsy}. 

%%%%%%%%%%%%%%%%%%%%%%%%%%%%%%%%%%%%%%%%
\subsection{Ansatz and boundary conditions}
%%%%%%%%%%%%%%%%%%%%%%%%%%%%%%%%%%%%%%
Imposing spherical symmetry, we employ for the scalar field the usual $O(3)$ Ansatz, which is parametrized in terms
of the radial function $f(r)$ and the harmonic time dependence 
\begin{equation}
    \phi^a=[\sin{f(r)}\cos{\omega t},~ \sin{f(r)}\sin{\omega t},~ \cos{f(r)}]
\label{Ansatz-phi}
\end{equation}
For the gauge potential we consider a purely electrostatic Ansatz
\begin{equation}
    A_\mu dx^\mu=A_0(r)dt \, .
\label{Ansatz-A}
\end{equation}

%%%%%%%%%%%%%%%%%%%%%%%%%%%%%%%%%%%%%%%%%%%%%%%%%%%%%%%%%%%%%%%%%%%
\begin{figure}[t!]
\begin{center}
\includegraphics[height=.285\textheight,  angle =-0]
{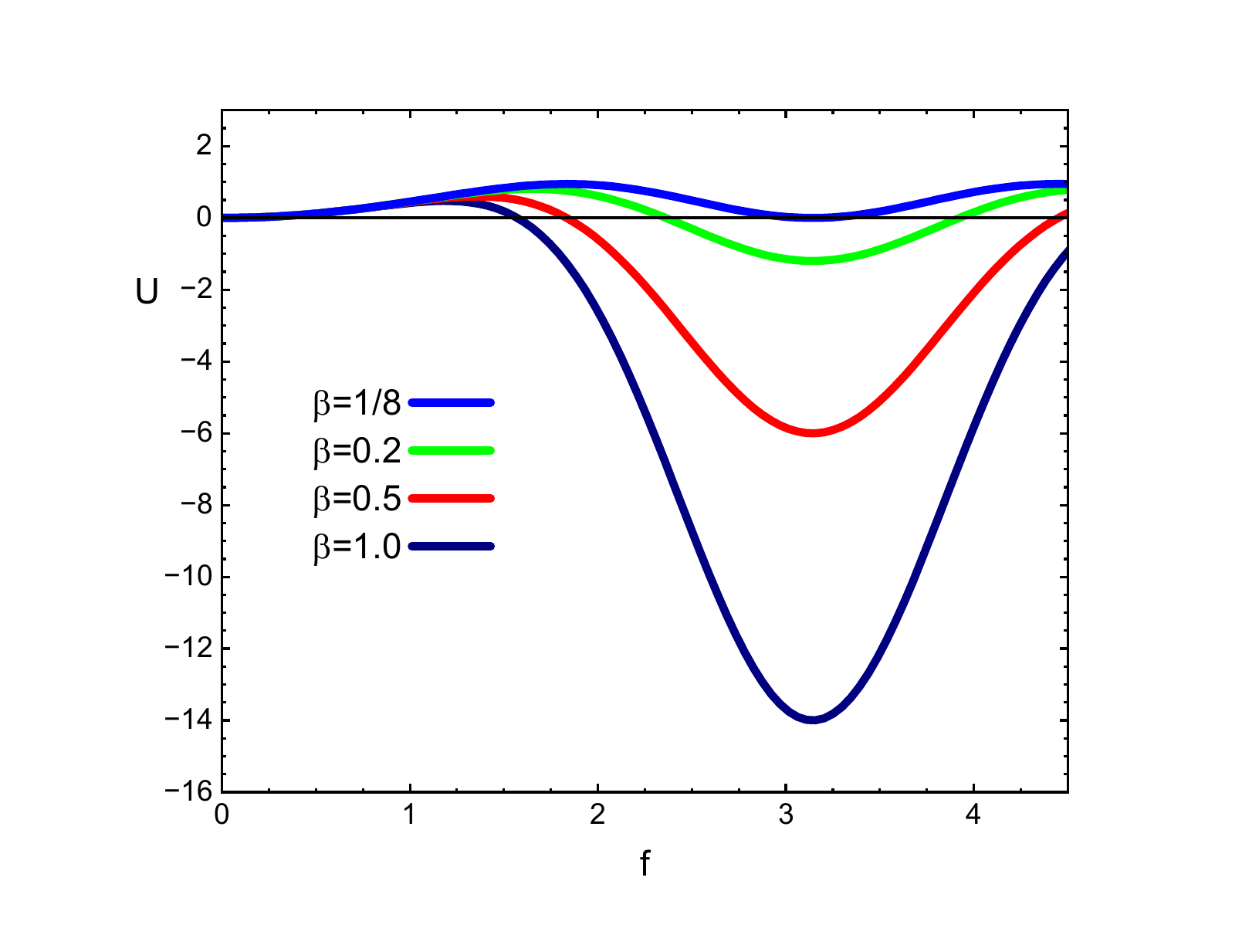}
\includegraphics[height=.285\textheight,  angle =-0]
{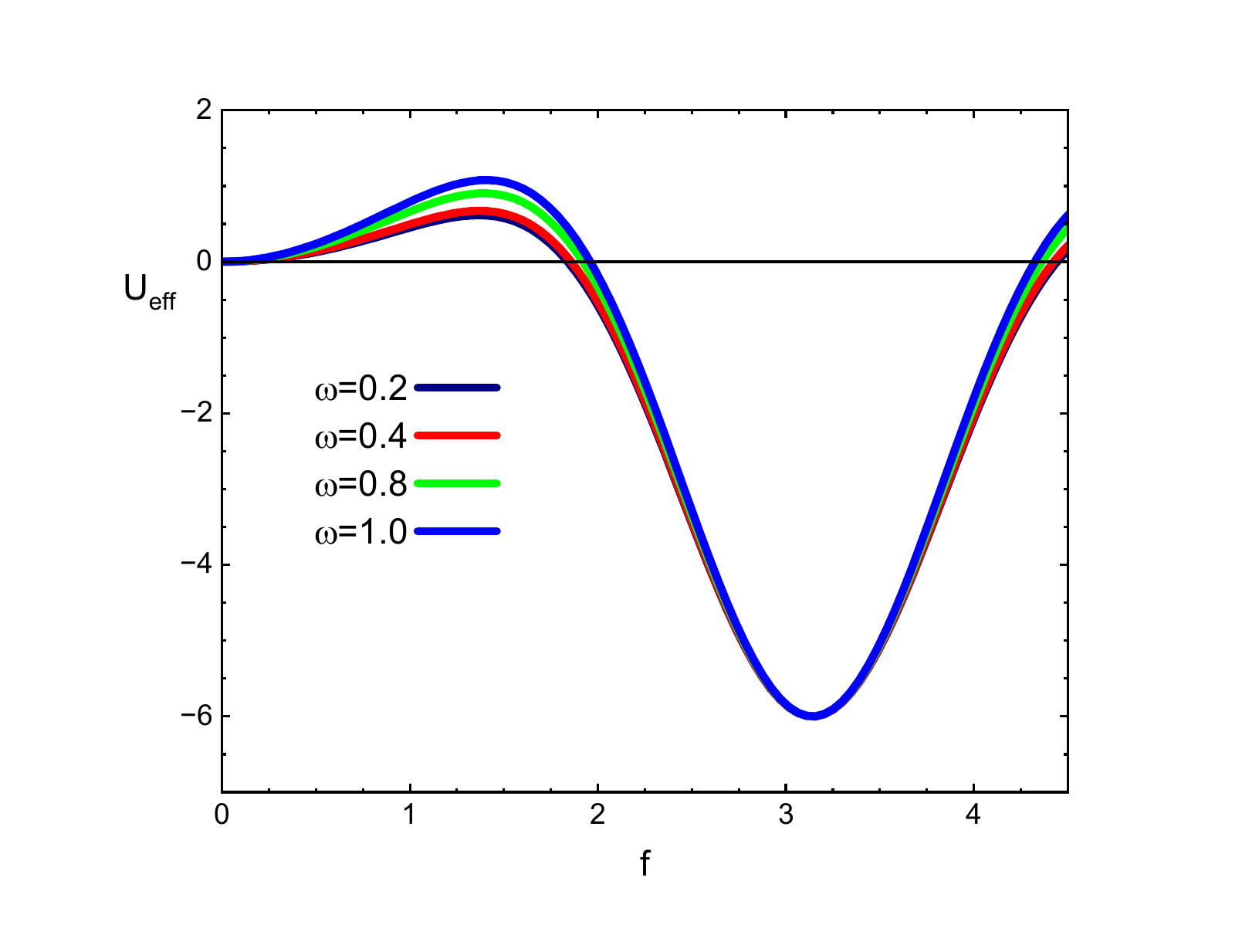}
\end{center}
\caption{\small The potential (\ref{pot}) for different values of the parameter $\beta$, $\mu^2=1$ (left) and the effective potential $U_{eff} = U+\frac12 \omega^2 \sin^2 f$ for $e=0$ and different values of the frequency $\omega$ at $\beta=0.5$ and $\mu^2=1$ (right), vs the value of the radial function $f$.}    
\lbfig{fig0u}
\end{figure}
%%%%%%%%%%%%%%%%%%%%%%%%%%%%%%%%%%%%%%%%%%%%%%%%%%%%%%%%%%%%%%%%%%

Note that the $U(1)$ invariance of the model \re{Gauged_O3} admits the static gauge fixing setting $\omega=0$. 
However, following our previous works \cite{Ferreira:2025xey,Kunz:2025tsy,Mikhaliuk:2025mxy}, we
employ the gauge fixing
$A_0\to 0$ as $r\to \infty$ and retain the dependence of the solutions on
the angular frequency $\omega$.

Substituting \re{Ansatz-phi},\re{Ansatz-A}  into \re{Gauged_O3}  we arrive at the
reduced Lagrangian
$$
L_{eff}=-\frac12 ( A_0^\prime)^2 -\frac12(\omega + eA_0)^2 \sin^2 f + \frac12 (f^\prime)^2 - \mu^2 (1 -\cos f) + \beta (1-\cos f)^4\,
$$
The corresponding field equations become
\be
\begin{split}
f^{\prime \prime} &+ \frac{2f^\prime}{r} + (\omega + eA_0)^2\sin f \cos f =
\left[\mu^2 - 4 \beta(1-\cos f)^3\right] \sin f,\\
A_0^{\prime \prime} &+
\frac{2A_0^\prime}{r}-e(\omega + eA_0)\sin^2 f =0\, ,
\end{split}
\label{sys-reduced}
\ee
where a prime denotes the radial derivative. 

As for the usual Q-balls, there are two important characteristics of the $U(1)$ gauged solitons of the $O(3)$ sigma model, the mass of the configurations
\be
M=\int d^3x\, T^{00}   =4\pi\!\int\! r^2dr \, {\left(\frac12 f'^2 + \frac12A_0'^2 + \frac12 (\omega + eA_0)^2 \sin^2{f} + \mu^2 (1\!-\!\cos f) \!-\! \beta(1\!-\!\cos f )^4 \right)}
\label{mass}
\ee
and the Noether charge
\be
Q=\int d^3x\, j^0=4\pi\!\int\! r^2dr \,{\sin^2{f}\left(\omega+eA_0\right)} \, .
\label{charge}
\ee

It must be emphasized that, unlike ordinary Q-balls in a complex scalar field theory,  the Noether charge $Q$ \re{charge} cannot be interpreted as the particle number since the norm of the $O(3)$ field is fixed, and only two components of the scalar triplet contribute to the
charge. On the other hand, the mass of the configuration \re{mass} includes the contributions of all components of the scalar field. Therefore, one can expect a different type of dependence $M(Q)$ for solitons of the $O(3)$ sigma model, and in the usual scalar theory.

It follows from the system of equations \re{sys-reduced} that, as the fields approach the vacuum values on the spacial asymptotics, 
$$f(r) \to 0 + h(r) + \dots,\quad A_0(r) \to  0 + a(r) + \dots \, ,
$$ the  linearized asymptotic equations describe perturbative excitations  around the vacuum:
\be
h^{\prime \prime} - \left(
\mu^2 -  \omega^2\right) h =0\, ,\qquad
a^{\prime \prime}  =0 \, . 
\label{decay-eqs}
\ee
Notably, the contribution of the term, which is proportional to $\beta$, decays faster than the usual mass term;  a localized configuration of the $O(3)$ field  with an exponentially decaying tail may
exist if $\omega \le \mu$ where $\mu$ is the mass of scalar perturbations. On the other hand, the gauge field remains massless.

%%%%%%%%%%%%%%%%%%%%%%%%%
\section{Mechanical properties of non-topological $O(3)$ solitons }
%%%%%%%%%%%%%%%%%%%%%%%%%
It was suggested by Maxim Polyakov \cite{Polyakov:2002yz,Polyakov:2018zvc} that
soliton-like configurations may be considered by analogy with an elastic medium. Explicitly, the spacial components of the stress-energy tensor \re{tmunu} can be associated
with the distribution of the pressure anisotropy (shear forces) $s(r)$ and the elastic
pressure $p(r)$  as

\begin{equation}
    T_{ij}= \left(\frac{x_i x_j}{r^2} - \frac13 \delta_{ij}  \right) s(r)
    + p(r)\delta_{ij}\, , \quad i,j=1,2,3 \, .
\end{equation}
Using the spherically symmetric Ansatz 
\re{Ansatz-phi},\re{Ansatz-A} and the expression for the energy-momentum tensor 
\re{tmunu}, we obtain
\be
\begin{split}
    p(r)&=-\frac16f'^2+\frac16A_0'^2+\frac12(\omega+eA_0)^2\sin^2{f}-\mu^2(1-\cos f) + \beta(1-\cos f)^4,
    \\
    s(r)&=-A_0'^2+f'^2,
\end{split}
\label{S_P}
\ee
Conservation of the energy-momentum tensor yields the 
\emph{equilibrium} equation \cite{Mai:2012yc,Loginov:2020xoj}
\be
d(r)=p^\prime(r)+\frac{2}{3}s^\prime(r)+\frac{2}{r}s(r)=0\, .
\ee
or
\be
p^\prime(r)=-\frac{2}{3r^3}\frac{\mathrm{d}}{\mathrm{d}r}\left [r^{3} s^2(r) \right ]
\label{stability-d}
\ee

Another criterion of stability of a soliton is 
the von Laue condition
\cite{Laue:1911lrk,Polyakov:2018zvc,Pinto:2025plg},
related to the repulsive and attractive forces balance in the interior of a soliton:
\be
\int\limits_0^\infty r^2 dr\, p(r) = 0
\ee
Indeed, integrating this equation by parts over $r$ and imposing a finite upper integration
limit $R$, we obtain
\be
\int\limits_0^R r^2 dr\, p(r) = \left[\frac{r^3}{3}p(r) \right]_0^R - \int \limits_0^R dr \frac{r^3}{3}p^\prime(r)
\ee
Using the relation \re{stability-d}, we obtain 
\be
\label{int-Laue}
\int\limits_0^R r^2 dr\, p(r) = \left[\frac{r^3}{3} \left(p(r) +\frac{2}{3}s(r)\right)\right]_0^R \, .  
\ee
Since the asymptotic behaviour  of the solutions of the system \re{sys-reduced} secures that 
the pressure and shear force densities decay at large distances faster than $1/r^3$, the von Laue condition is satisfied.
Consequently, the pressure function $p(r)$ must possess at least one zero.
This is a necessary (though not sufficient)
condition of stability; it can also be reformulated as a virial relation connecting
integrals of components of the energy-momentum tensor. 

The expression in brackets in \re{int-Laue} corresponds to
the distribution of the normal component of the net force
acting on an infinitesimal area element $dA\, \boldsymbol{\hat{e}}_{r}$ at a distance $r$
\cite{Polyakov:2018zvc,Perevalova:2016dln}, $dF_{j}=T_{jk}\, dA_{k}$ and the normal force must be directed outward
\be
c(r)=\frac{dF_{r}}{dA_{r}}=p(r) +\frac{2}{3}s(r) > 0
\label{C-criterion}
\ee

Further, by analogy with related studies of the
internal structure of hadrons, one can introduce one more
mechanical characteristic, the so-called \emph{D-term} (or the \emph{Druck-term}) which is related to
the distribution of the internal forces inside the configuration \cite{Polyakov:2002yz}.
It can be expressed in terms of the pressure and shear force distributions as:
\be
    D=-\frac{M}{3}\!\int d^3x~r^2 s(r)=\frac54 M\!\int d^3x~r^2p(r),
\ee
Indeed, there is yet another relation between the energy density $\epsilon(r)=T_{00}(r)$, the pressure $p(r)$, shear forces $s(r)$ and the charge density $j^0(r)$:
\be
    \epsilon+p=(\omega +eA_0)j^0+\frac13 s+A_0'^2,
\ee
Multiplying this relation by $r^2$ and integrating over $d^3x$, one can express $D$ in terms of other properties:
\be
    D=\frac{5}{9}\left(\omega MQ<r_Q^2>-M^2<r_E^2>\right),
    \label{d-term-from-prop}
\ee
where the mean square radii of the energy and charge densities are:
\be
   <r_E^2>=\frac{\int T_{00}r^2~d^3x}{\int T_{00}~d^3x}, \quad
   <r_Q^2>=\frac{\int j_{0}r^2~d^3x}{\int j_{0}~d^3x},
\ee
respectively. It was suggested that the condition for the stability of the configuration is a negative value of the D-term \cite{Polyakov:2018zvc}. 

Finally, localized configurations usually satisfy the energy conditions. 
They follow from the assumption that the energy density should
be positive everywhere in space. Explicitly, the following energy conditions must hold:
\begin{align}
    {\rm Weak~~energy~~condition:}~~~ &\varepsilon=T_{\mu \nu}X^{\mu}X^{\nu}\ge0,
    \label{weak}\\
{\rm Strong~~energy~~condition:}~~~    &\rho=(T_{\mu\nu}-\frac{T}{2}\eta_{\mu\nu})X^\mu X^\nu\ge 0,
\label{strong}
\end{align}
where $X^\mu = (1,0,0,0)$ is a timelike vector field. As we will see, both conditions can be violated for the soliton solutions of the model \re{Gauged_O3}.

\section{Numerical results}
The system \re{sys-reduced} can be solved numerically.
The corresponding boundary conditions are found by
considering the asymptotic expansions of the equations
on the boundary of the domain of integration together with
the assumption of regularity.
Explicitly, for the spherically symmetric configurations we impose
$A_0= f=0$ at spacial infinity and
$\partial_r A_0=\partial_r f=0$ at the origin.  

Numerical calculations are performed on an equidistant one-dimensional grid, employing
the compact radial coordinate $x=r/(C+r) \in [0:1] $ where $C$ is an arbitrary constant used to
adjust the grid according to the contraction of the solutions. In our numerical calculations we have
made use of a sixth-order finite difference scheme, where
the system of equations is discretized on a grid with a typical size of about 200 points in the radial direction. The emerging system of nonlinear algebraic equations has been solved
using the Newton-Raphson scheme, typical errors are of order $10^{-8}$.   

Let us briefly review the basic properties  of the spherically symmetric solutions of the model \re{Gauged_O3}  discussed previously in \cite{Ferreira:2025xey}. 

%%%%%%%%%%%%%%%%%%%%%%%%%%%%%%%%%%%%%%%%%%%%%%%%%%%%%%%%%%%%%%%%%%%%
\begin{figure}[t!]
\begin{center}
\includegraphics[height=.285\textheight,  angle =-0]
{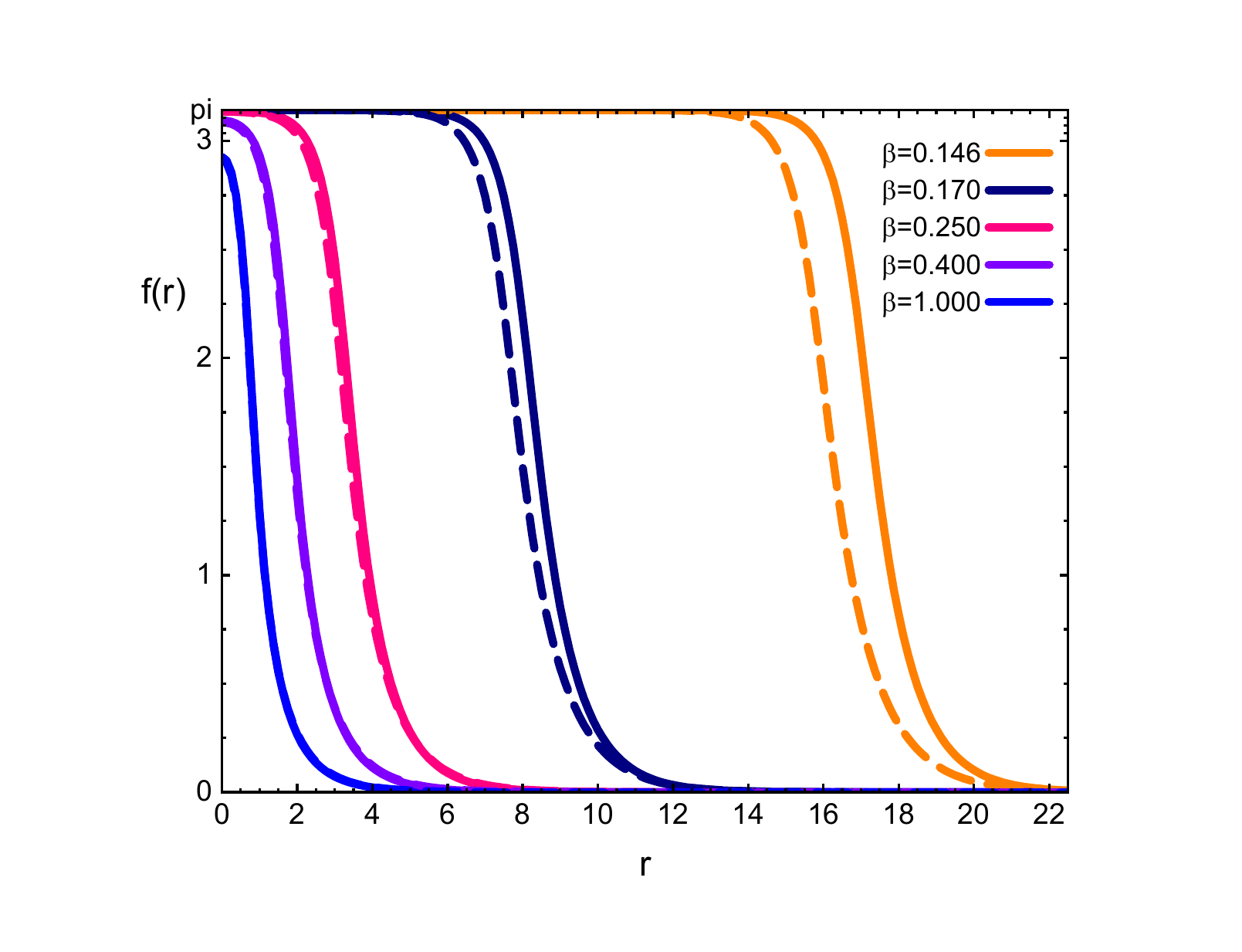}
\includegraphics[height=.285\textheight,  angle =-0]
{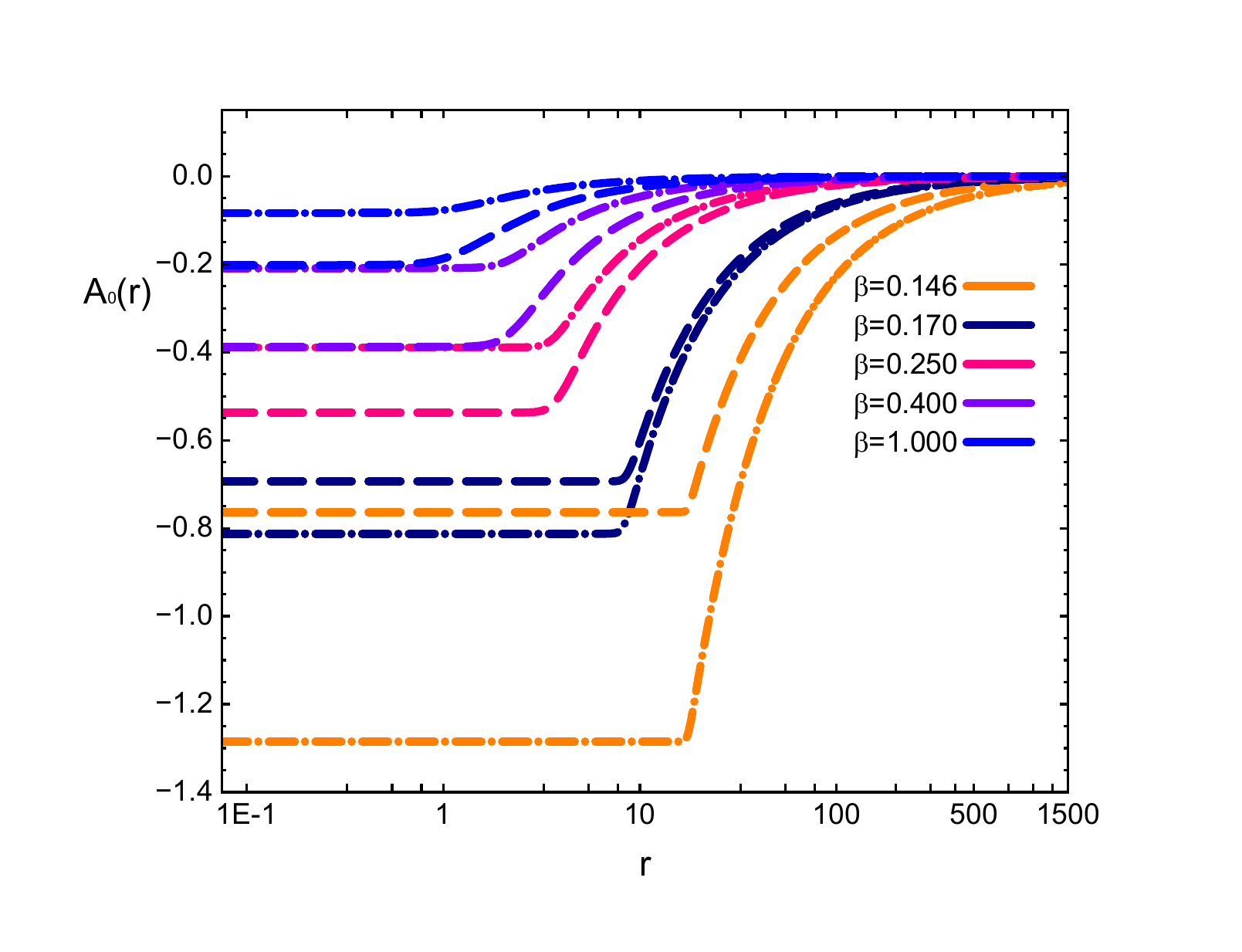}
\end{center}
\caption{\small Profiles of the radial profile functions of the soliton $f(r),A_0(r)$ for different values of the parameter $\beta$ at $\omega=0.5$ and $e=0$ (solid lines), $e=0.2$ (dashed doted lines) and $e=0.6$ (dashed lines). (A logarithmic scale is used for the function $A_0$).} 
\lbfig{fig1a}
\end{figure}
%%%%%%%%%%%%%%%%%%%%%%%%%%%%%%%%%%%%%%%%%%%%%%%%%%%%%%%%%%%%%%%%%%%%

In Fig. \ref{fig1a} we present radial profiles of the scalar and gauge fields for a set of values of the parameter $\beta$ and the gauge coupling $e$. For a fixed value of $e$, the size of the configuration rapidly increases as  $\beta$ decreases towards the critical value $\beta_c=1/8$. An increase of the gauge coupling $e$ leads to the opposite effect, as seen in Fig.~\ref{fig1a},  left plot. The central value of the profile function of the scalar field approaches \textit{anti-vacuum} $\phi^3 =(0,0-1)$; it decreases as $\beta$ grows, the short-range attractive force becomes stronger and the core of the soliton shrinks.    

The electric potential $A_0$ is a
constant in the interior of the soliton, as displayed in Fig.~\ref{fig1a}, right plot. The electric field of the gauged configuration vanishes in the interior. Outside the
core of the soliton, it possesses a long-range Coulomb asymptotic
tail sourced by the charge $Q_e=eQ$. 

%%%%%%%%%%%%%%%%%%%%%%%%%%%%%%%%%%%%%%%%%%%%%%%%%%%%%%%%%%%%%%%%%%%
\begin{figure}[t!]
\begin{center}
\includegraphics[height=.285\textheight,  angle =-0]{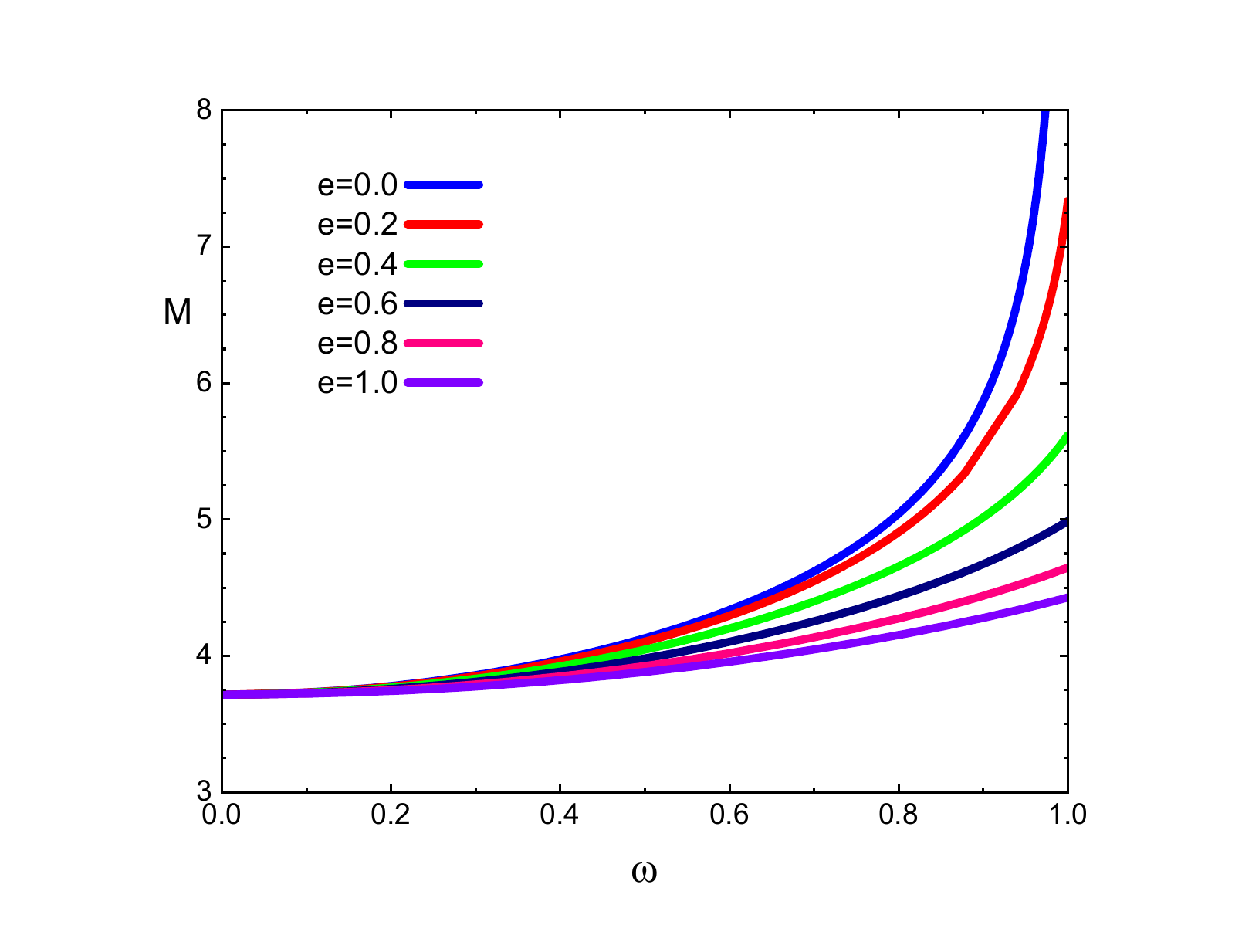}
\includegraphics[height=.285\textheight,  angle =-0]{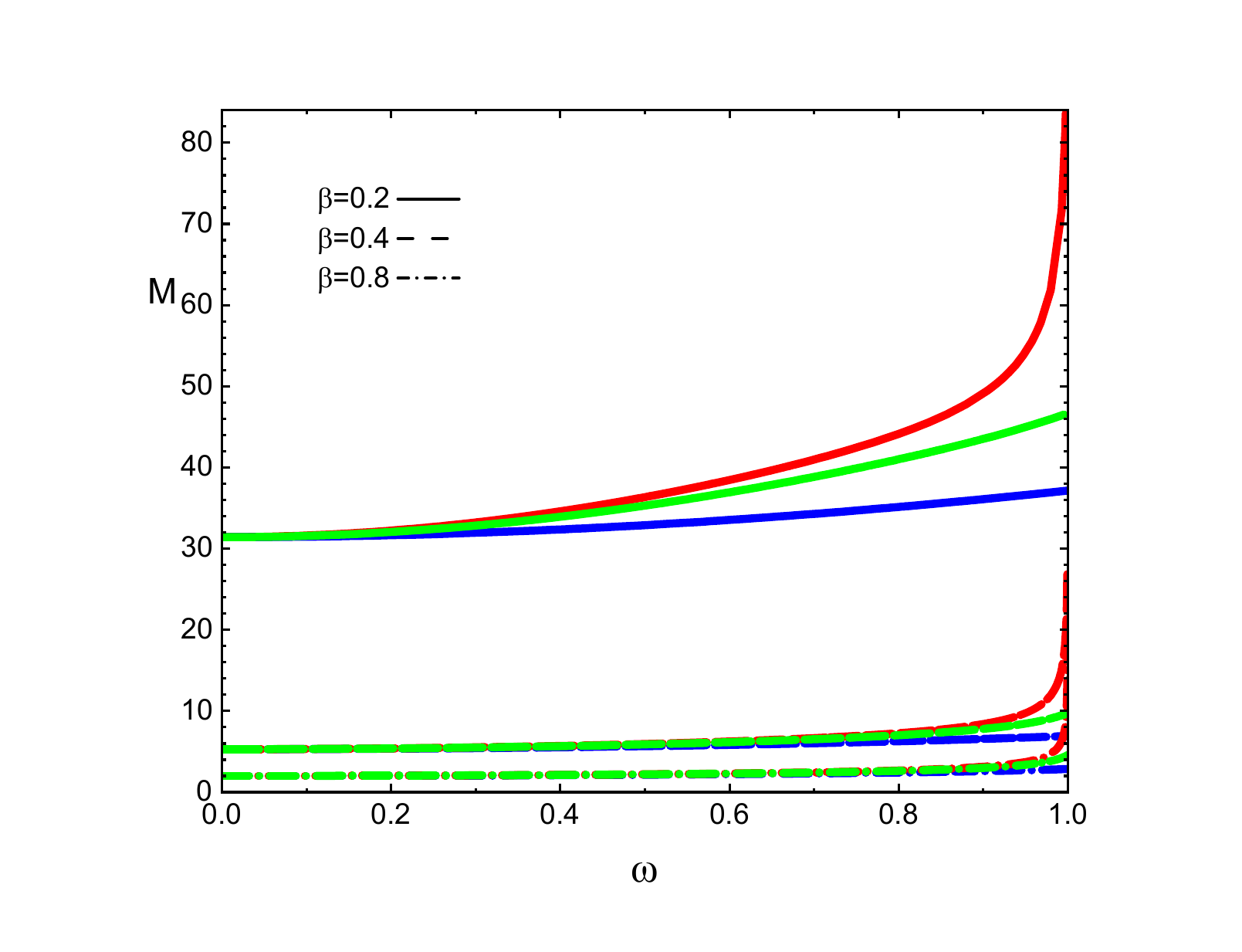}
\includegraphics[height=.285\textheight,  angle =-0]{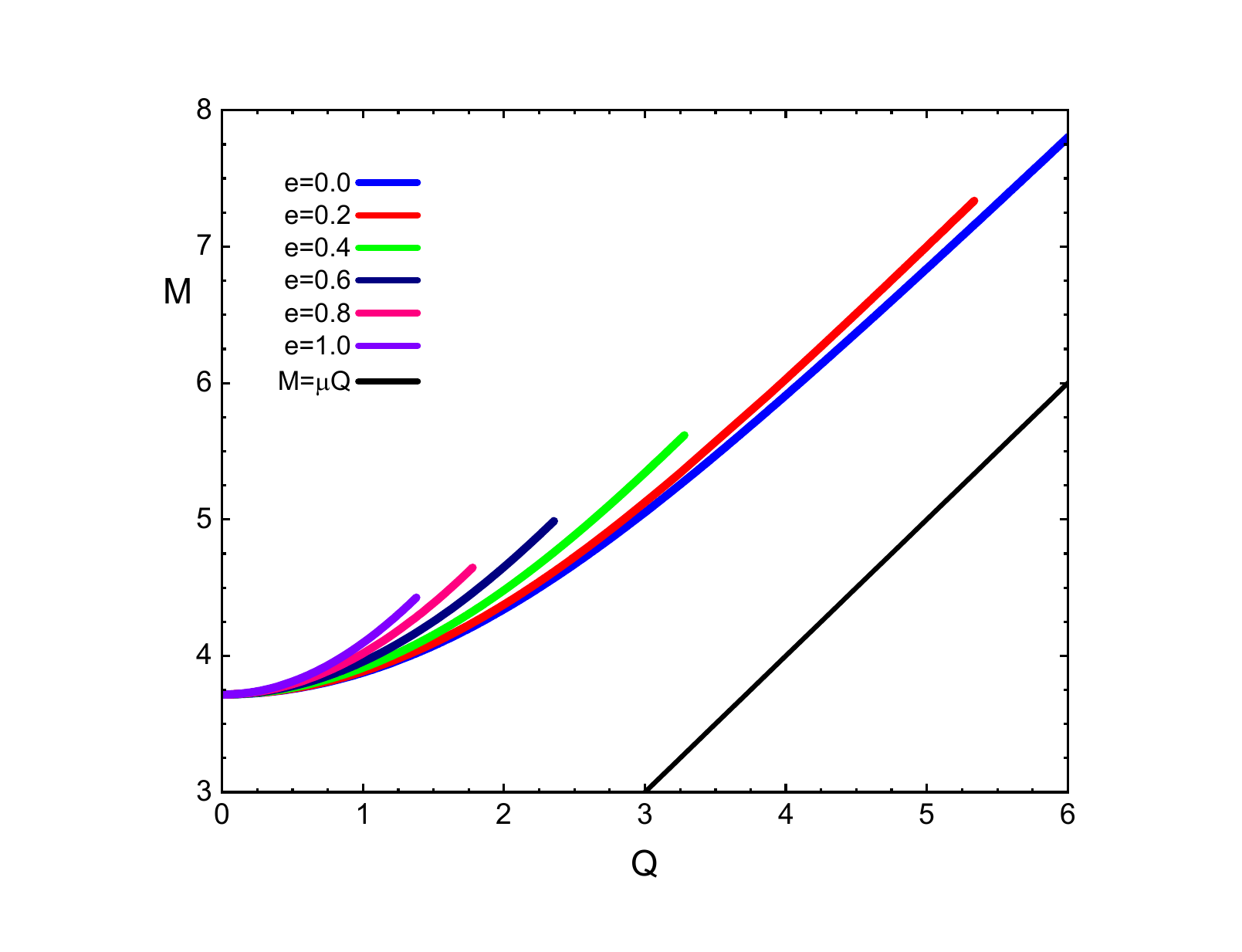}
\includegraphics[height=.285\textheight,  angle =-0]{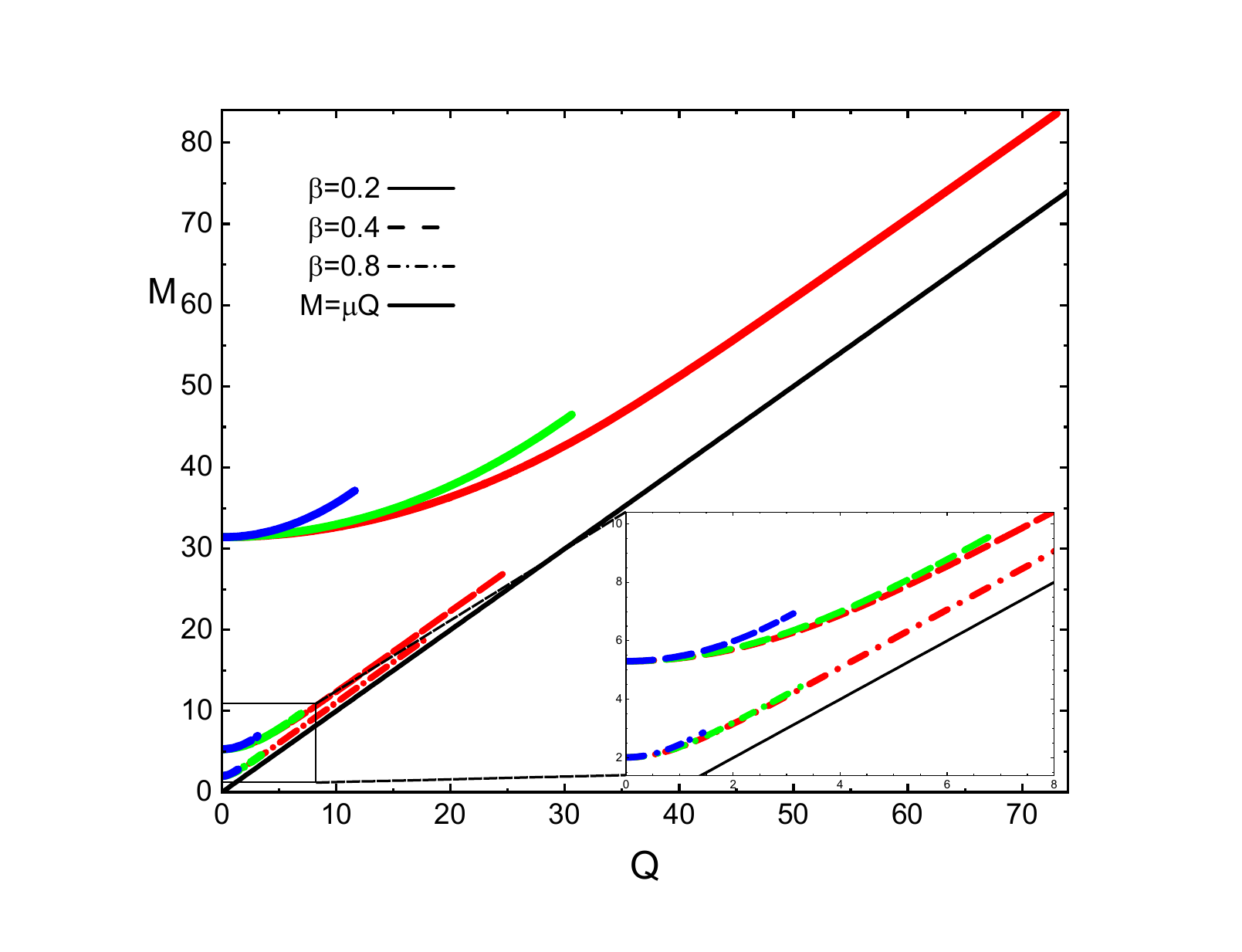}
\end{center}
\caption{\small Spherically symmetric $U(1)$ gauged $O(3)$ solitons: 
The mass $M$ vs the frequency $\omega$ (upper plots) and vs the Noether charge $Q$ (bottom plots) for some set of values of the gauge coupling $e$ for $\beta=0.5$ (left) and for some set of values of the parameter $\beta$ (right) for $e=0$ (red line), $e=0.2$ (green line), $e=0.6$ (blue line).}    
\lbfig{fig1}
\end{figure}
%%%%%%%%%%%%%%%%%%%%%%%%%%%%%%%%%%%%%%%%%%%%%%%%%%%%%%%%%%%%%%%%%%%

Soliton solutions of the model \re{Gauged_O3} exist for all nonzero values of the  frequency $\omega \in [0,1]$ where
the upper critical value corresponds to the mass of free charged quanta of scalar excitations. 

It is worth emphasizing the following important difference between the solutions under consideration and ordinary Q-balls in the theory of a complex scalar field. In the latter case Q-balls arise smoothly from perturbative excitations about
the vacuum, as the angular frequency is decreasing below the mass threshold, the  scalar quanta condense into a non-topological soliton. The situation is different  for the solitonic solutions of the $O(3)$ sigma-model with potential \re{pot}, they
are disconnected from the vacuum excitations for all the range of values of the frequency $\omega$ and  the parameter $\beta$. 

In Fig.~\ref{fig1}  we present the total mass $M$ of the
solutions versus the angular frequency $\omega$ (upper plots) and Noether charge $Q$ (bottom plots) for a set of values of the
parameter $\beta$ and fixed gauge coupling $e$ and vice versa. 

The gauged solitons form a branch, which extends forwards as $\omega$ increases. The limiting fundamental
solution at $\omega=0$ is uncharged; it corresponds to the
non-topological soliton stabilized by potential \re{pot}. 

The mass of the configuration increases monotonically with increasing frequency, in the case of ungauged solitons ($e=0$) it diverges as the angular frequency approaches the upper critical value, as seen in  Fig.~\ref{fig1}, upper plots. We observe that the critical behaviour of the electrically charged solitons is different, they possess
finite energy and charge for all ranges of values of the angular frequency and parameter $\beta$, see Fig.~\ref{fig1}, upper plots. Increasing of the gauge coupling $e$  leads to a decrease of the mass while, at fixed charge $Q$, the mass $M$ increases with increasing coupling $e$. This behavior can be attributed to the simultaneous enhancement of the electrostatic repulsion and the attractive contribution associated with the structure of the potential \re{pot}.

The parameter $\beta$ has a crucial impact on both the solutions and the mass. As $\beta$ approaches the critical value $\beta_c=\frac{1}{8}$, the mass of the configuration rapidly increases. On the other hand, for a fixed value of $e$, the mass–charge relation approaches the linear form $M = \mu Q$, as $\beta$ increases, see Fig.~\ref{fig1}, bottom right plot. 

\begin{figure}[t!]
\begin{center}
\includegraphics[height=.285\textheight,  angle =-0]
{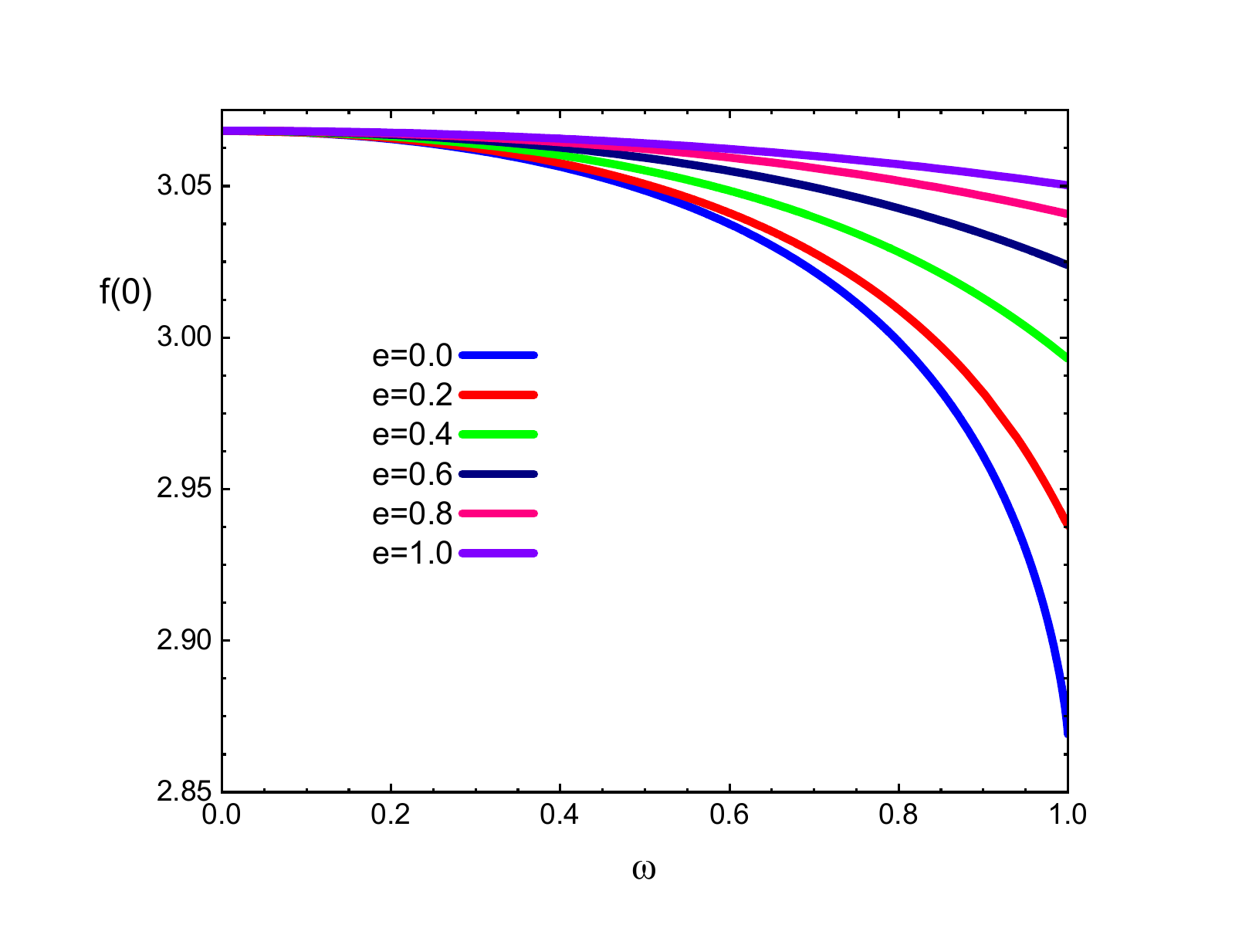}
\includegraphics[height=.285\textheight,  angle =-0]
{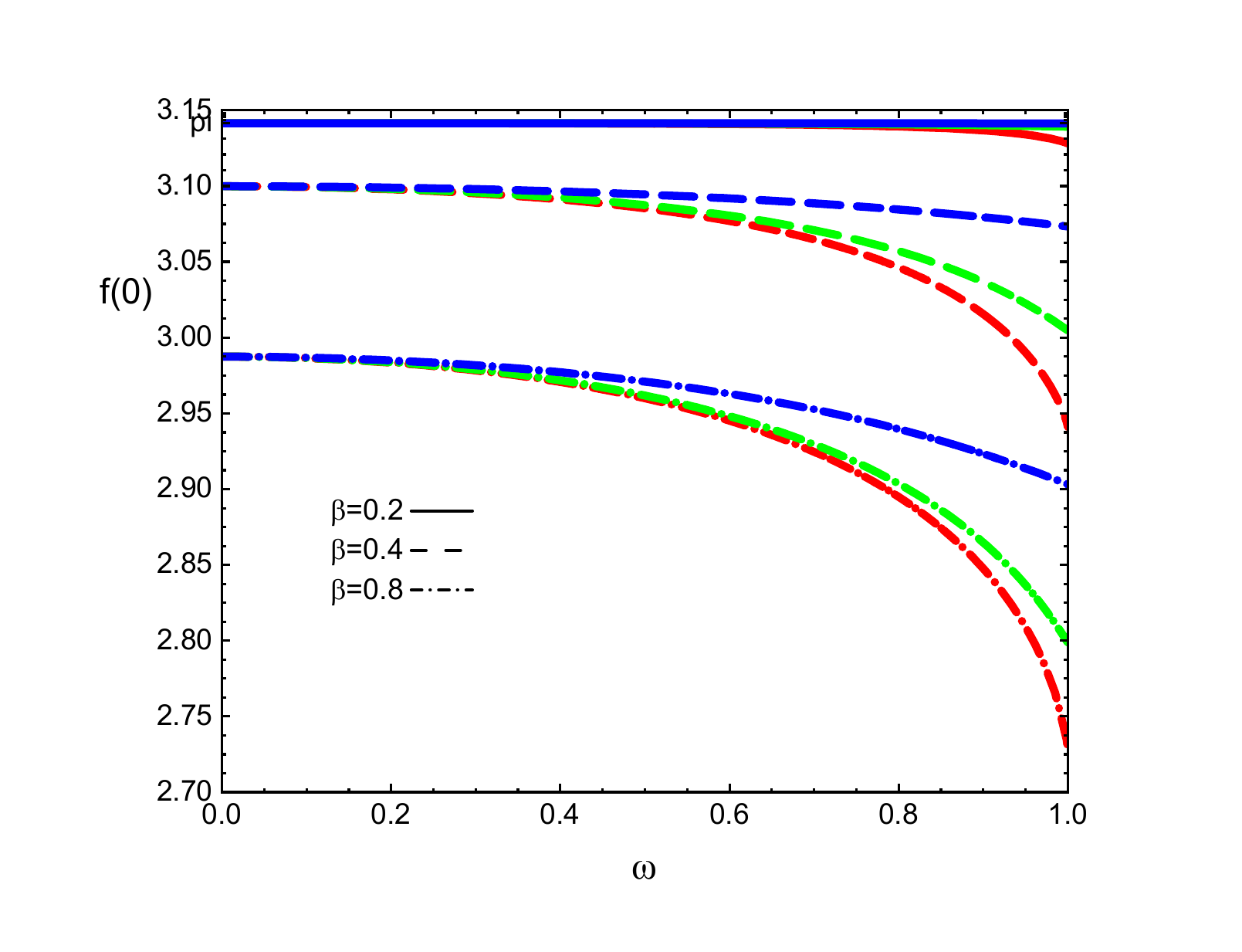}
\end{center}
\caption{\small The central value of the scalar profile function $f(0)$ vs the frequency $\omega$ for some set of values of $e$ at $\beta=0.5$ (left plot) and for some set of values of $\beta$ at $e=0$ (red line), $e=0.2$ (green line) and $e=0.6$ (blue line), right plot.}    
\lbfig{fig5}
\end{figure}
%%%%%%%%%%%%%%%%%%%%%%%%%%%%%%%%%%%%%%%%%%%%%%%%%%%%%%%%%%%%%%%%%%%%

As can be seen from Fig.~\ref{fig5}, the central value of the scalar profile function $f(0)$ decreases as the frequency $\omega$ increases all the way up to the mass threshold. The decrease rate is slower for electrically charged solitons, also the value of the profile function at the center of a soliton decreases faster for larger values of $\beta$.    

%%%%%%%%%%%%%%%%%%%%%%%%%%%%%%%%%%%%%%%%%%%%%%%%%%%%%%%%%%%%%%%%%%%
\begin{figure}[t!]
\begin{center}
\includegraphics[height=.285\textheight,  angle =-0]
{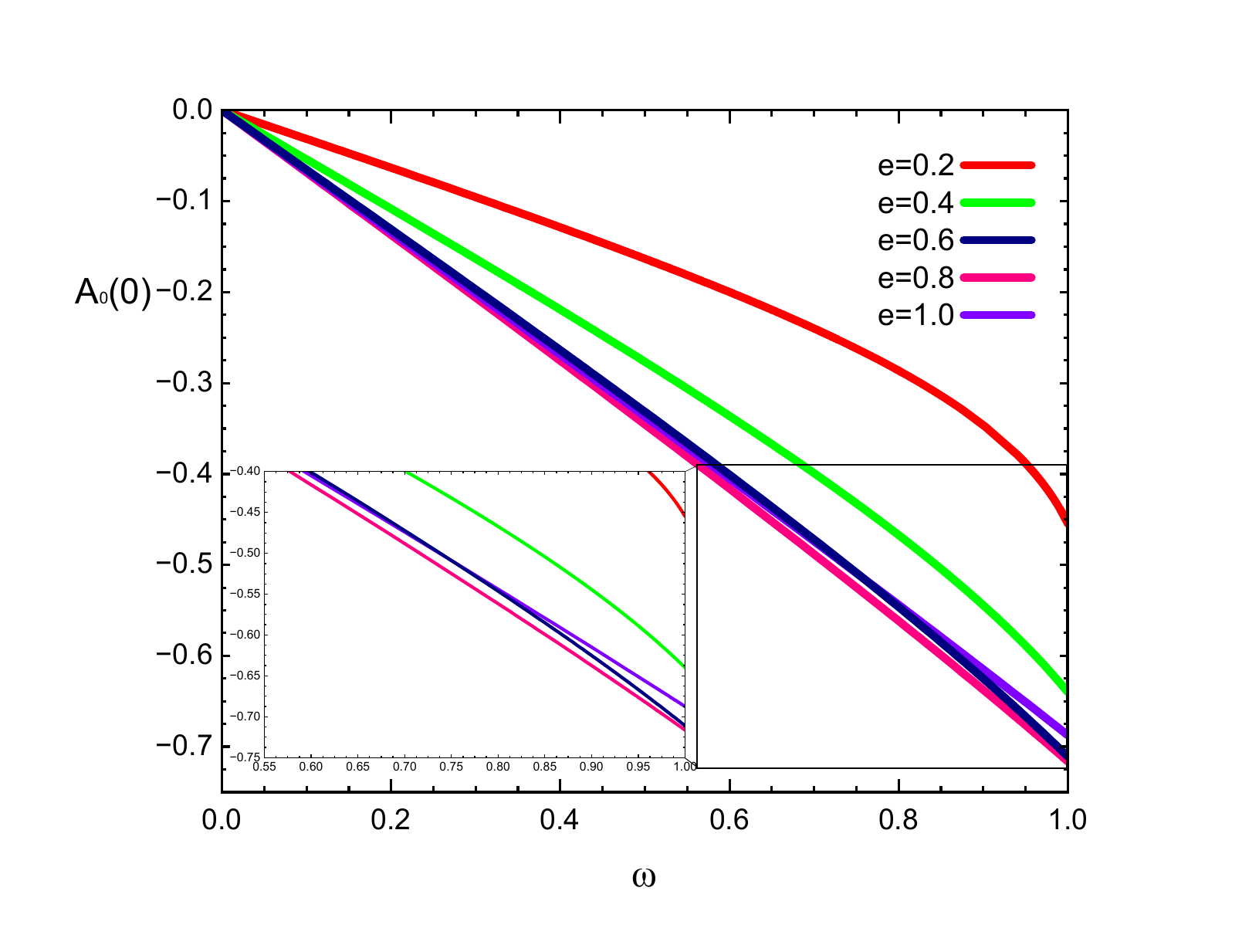}
\includegraphics[height=.285\textheight,  angle =-0]
{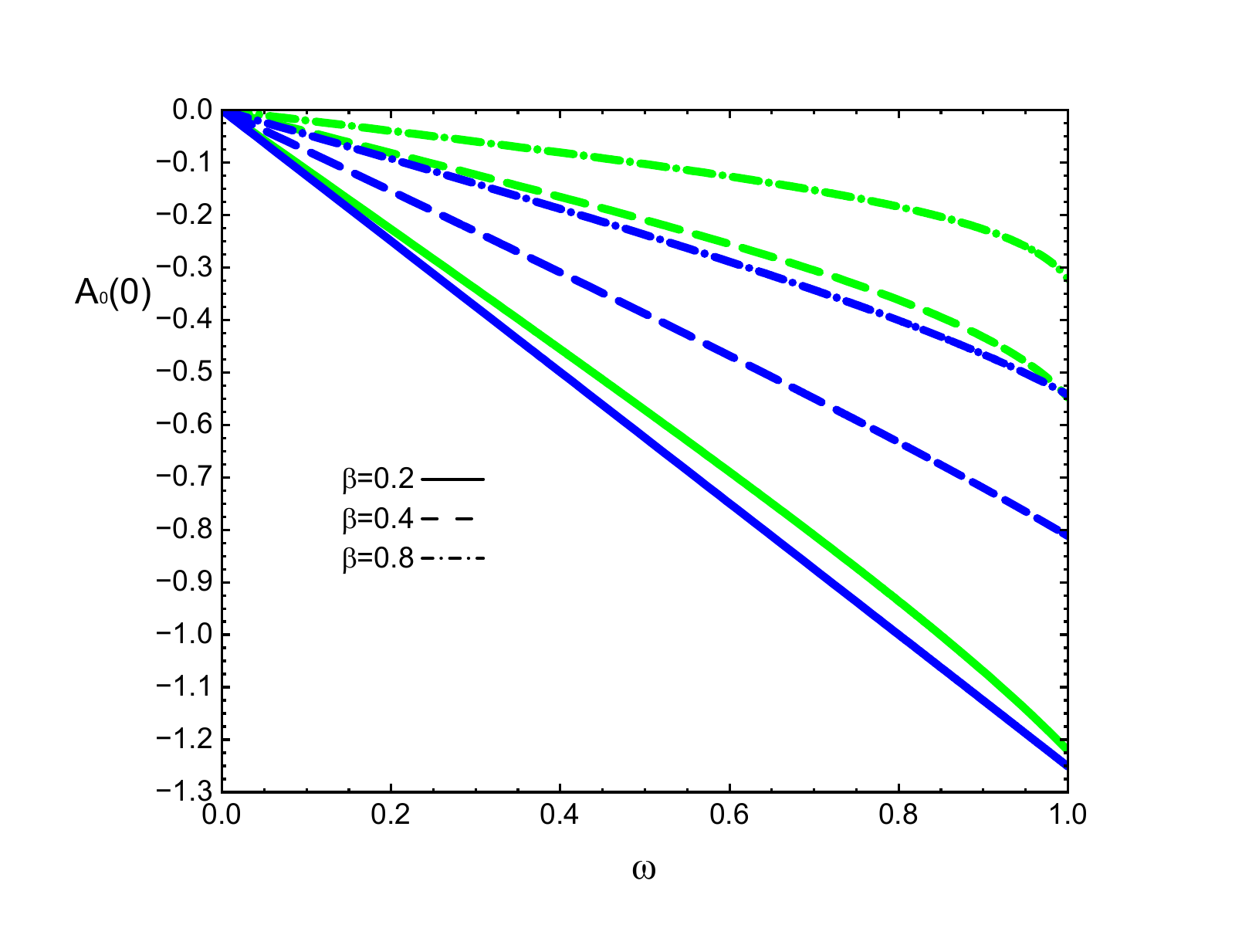}
\end{center}
\caption{\small The central value of electric potential $A_0(0)$ vs the frequency $\omega$ for some set of values of $e$ at $\beta=0.5$ (left plot) and for some set of values of $\beta$ at $e=0.2$ (green line) and $e=0.6$ (blue line), right plot.}    
\lbfig{fig7}
\end{figure}
%%%%%%%%%%%%%%%%%%%%%%%%%%%%%%%%%%%%%%%%%%%%%%%%%%%%%%%%%%%%%%%%%%%
In Fig.~\ref{fig7} we display the central value of electric potential $A_0(0)$ vs frequency $\omega$ for some set of values of the gauge coupling $e$ and the parameter $\beta$. This quantity indicates the strength of electrostatic repulsion. Clearly, for a given $\beta$, it decreases monotonically as $\omega$ increases. The increase of $\beta$ results in slower decay rate. 
Due to the interplay of two compensating forces of electrostatic repulsion and scalar interaction, the function $A_0(0)$ exhibits a nontrivial dependence on $e$, in particular, for $e>0.8$, the slope of the $A_0(0)$ vs. $\omega$ curve stops increasing, see Fig.~\ref{fig7}  left plot.

%%%%%%%%%%%%%%%%%%%%%%%%%%%%%%%%%%%%%%%%%%%%%%%%%%%%%%%%%%%%%%%%%%%
\begin{figure}[h!]
\begin{center}
\includegraphics[height=.285\textheight,  angle =-0]
{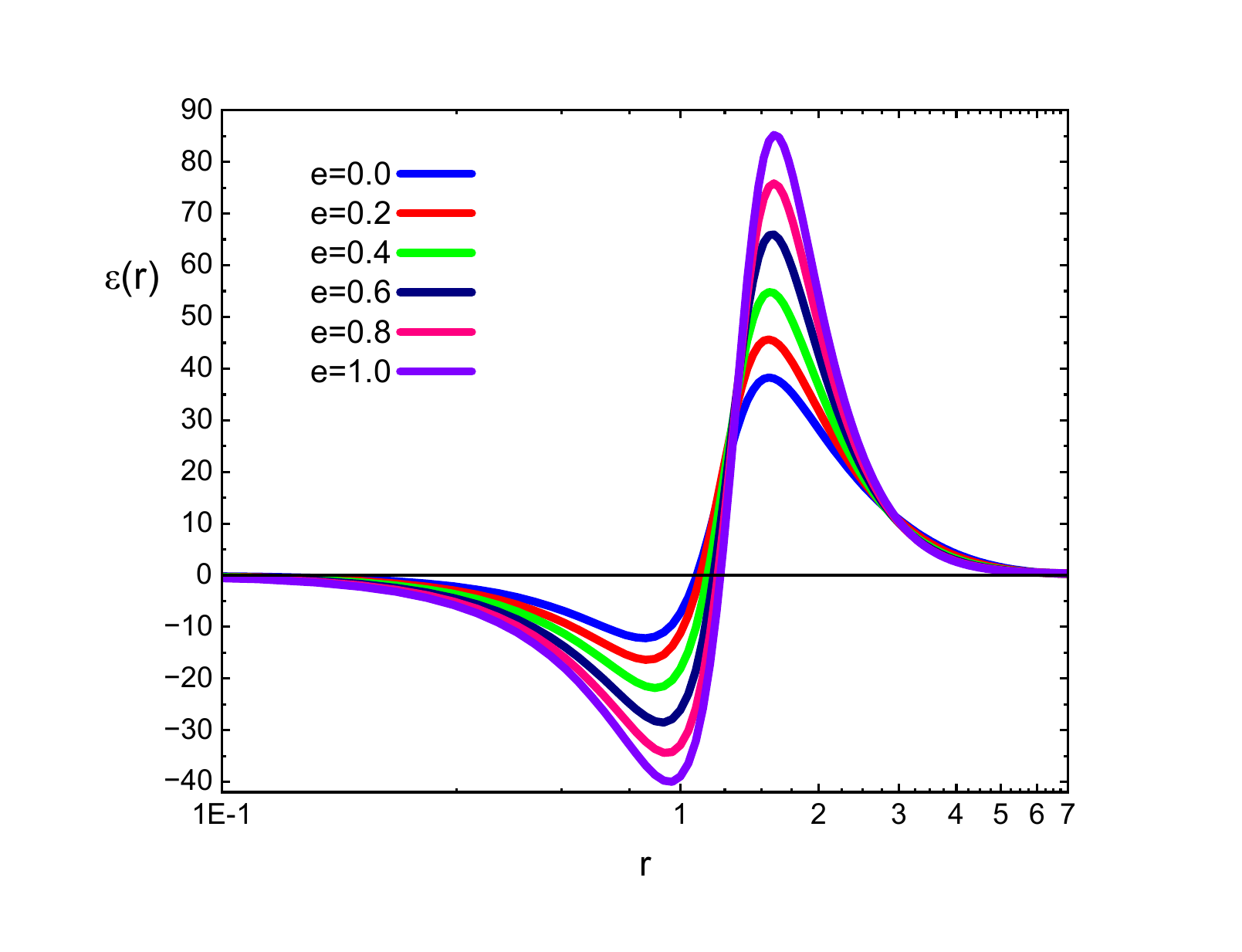}
\includegraphics[height=.285\textheight,  angle =-0]
{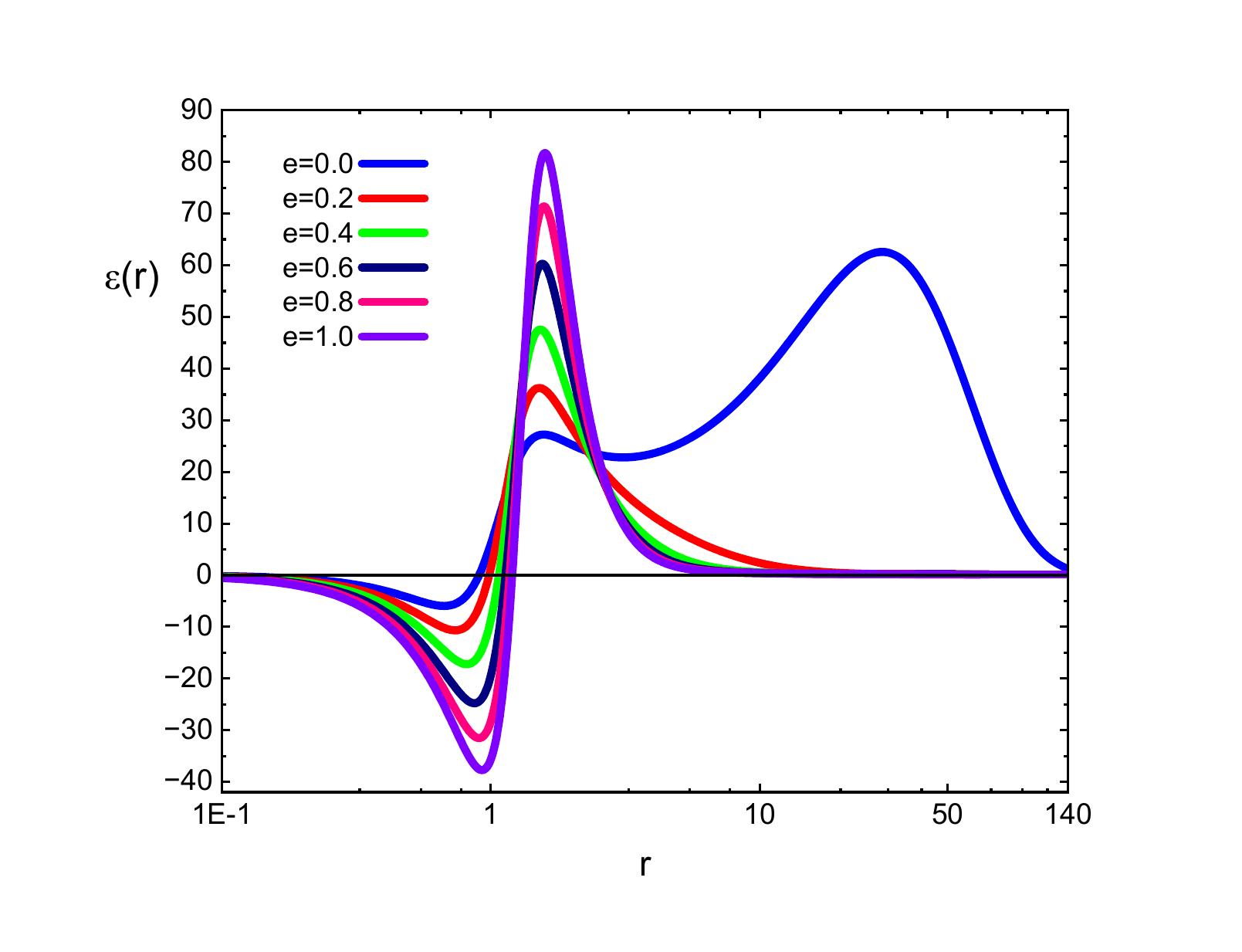}
\includegraphics[height=.285\textheight,  angle =-0]
{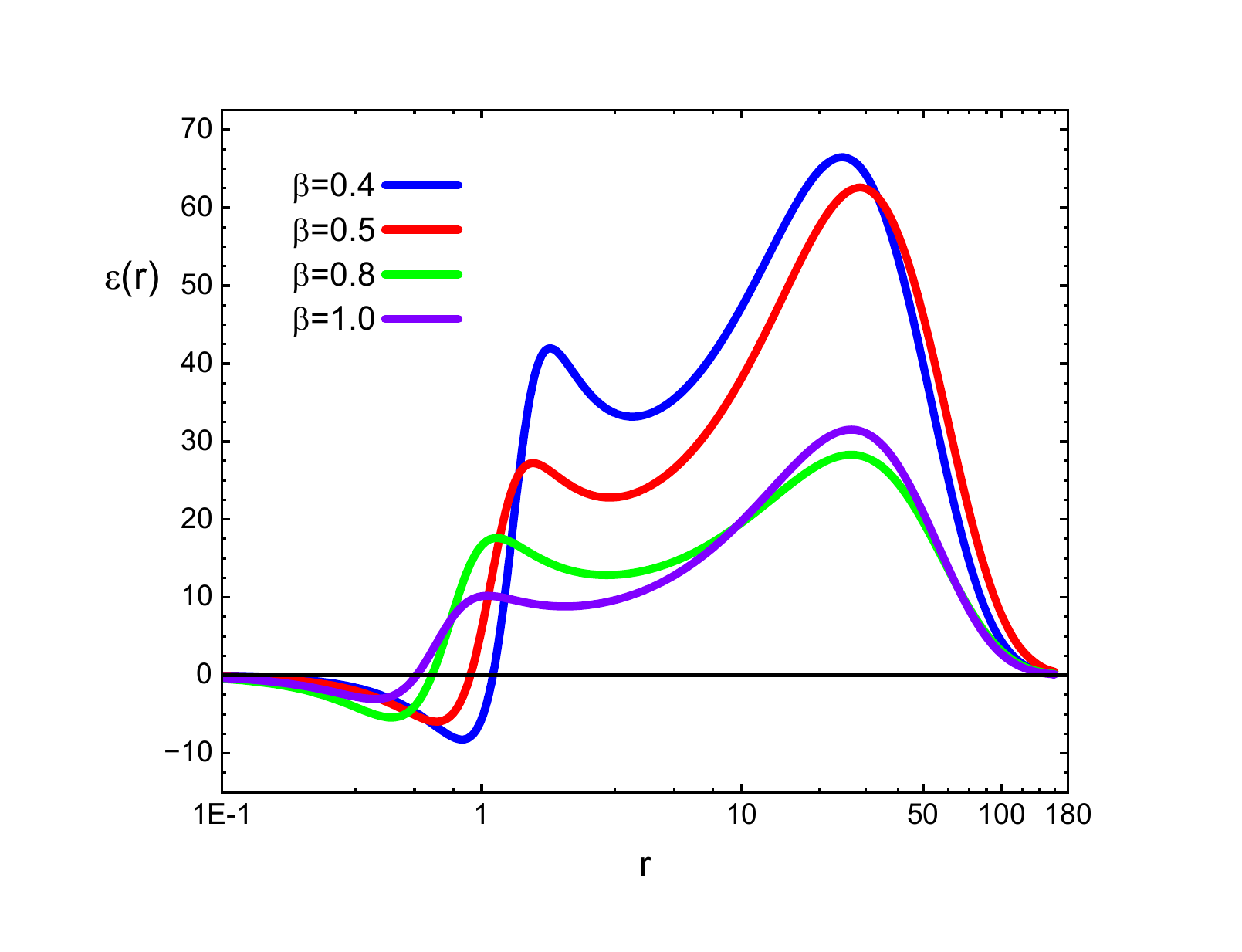}
\includegraphics[height=.285\textheight,  angle =-0]
{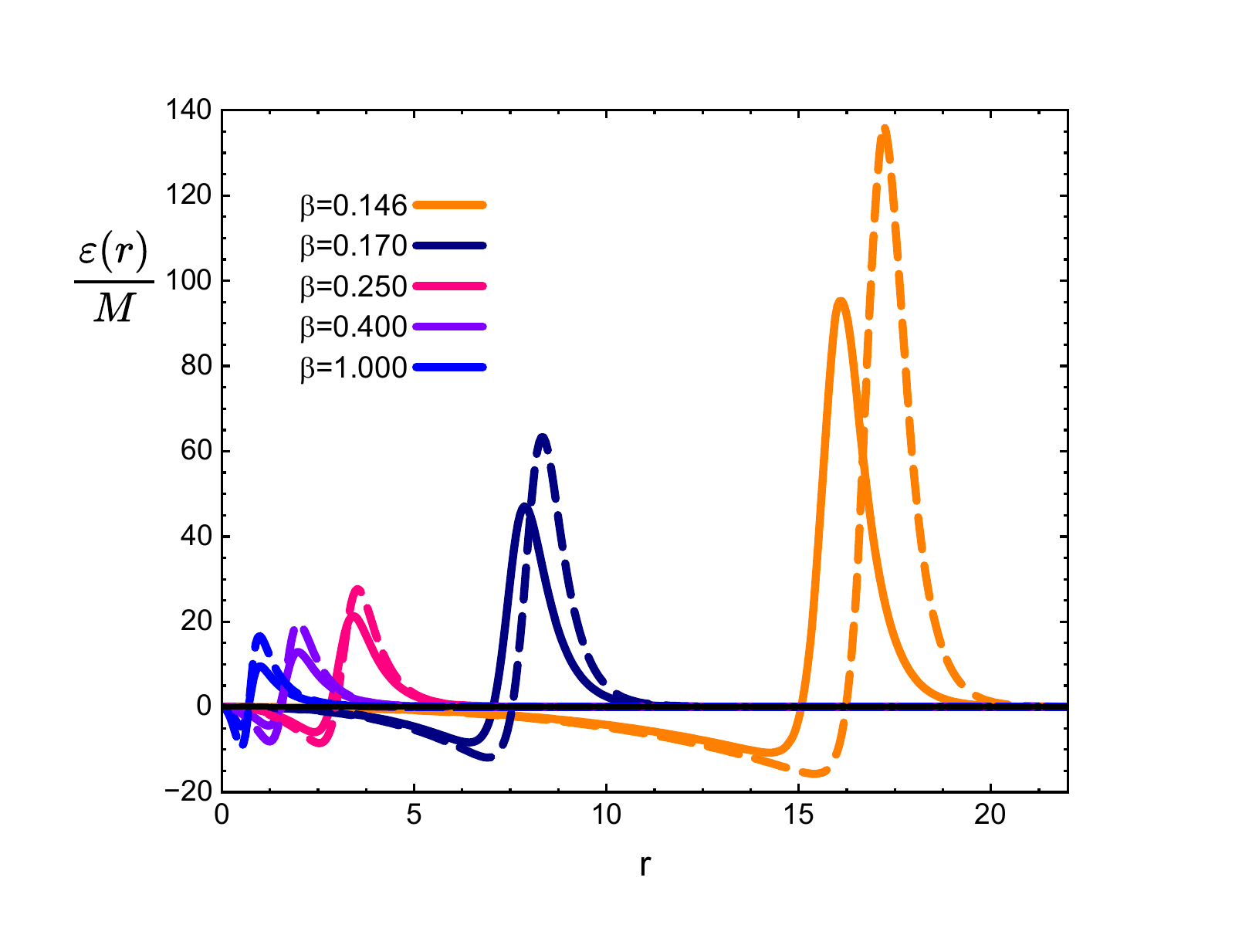}
\end{center}
\caption{\small Upper plots: Radial profiles of the energy density distributions $\varepsilon(r)$ are displayed for some set of values of $e$  for $\beta=0.5$ at $\omega=0.8$ (upper left) and $\omega=0.995$ 
(upper right). Bottom plots: Radial profiles of the energy density distributions $\varepsilon(r)$ are displayed for some set of values of $\beta$ at $e=0$ and  
$\omega=0.995$ (bottom left), and at $e=0$ (solid line), $e=0.6$ (dashed line) for $\omega=0.5$ (bottom right, normalized by $M$).}   
\lbfig{fig2}
\end{figure}
%%%%%%%%%%%%%%%%%%%%%%%%%%%%%%%%%%%%%%%%%%%%%%%%%%%%%%%%%%%%%%%%%%%%
Remarkably, for all solitons, the energy density distribution is not positive everywhere, see Fig.~\ref{fig2}, although the total integrated mass remains positive in any case. The structure of the potential \re{pot} leads to the appearance of a domain with negative energy density in the interior of the soliton. Hence, the weak energy condition  
\re{weak} is always violated, as expected. 
For a given value of the parameter $\beta$ and a fixed frequency $\omega$, an increase in the constant $e$ leads to an increase 
in both domains of positive and negative energy densities, as displayed in Fig.~\ref{fig2}, upper plots. Note that, for ungauged solitons, in the non-relativistic limit, as the frequency approaches the mass threshold, the function of the radial energy distribution ceases to be monotonic in the positive domain, see Fig.~\ref{fig2}.  For example, as $\omega = 0.995$ and $e = 0.0$, the function $\varepsilon(r)$ develops a pronounced hump and becomes more extended, approaching zero only at $r = 140$, see Fig.~\ref{fig2}, bottom left plot. \par

In Fig.~\ref{fig4a} we exhibit radial profiles of the function $\rho(r)$, which represents the strong energy condition~\re{strong}. As it turns out, this condition holds for some set of values of parameters of the model. For example, this condition is satisfied for the ungauged configurations at $\beta=0.5$, see Fig.~\ref{fig4a} upper plots. However, it becomes violated as the gauge coupling increases.  For $\omega<\omega_c=0.8$, the strong energy condition is violated for all values of $e$ while $\beta=0.5$. The critical value of the frequency increases with increasing $e$ and decreasing $\beta$.

%%%%%%%%%%%%%%%%%%%%%%%%%%%%%%%%%%%%%%%%%%%%%%%%%%%%%%%%%%%%%%%%%%%%
\begin{figure}[t!]
\begin{center}
\includegraphics[height=.285\textheight,  angle =-0]
{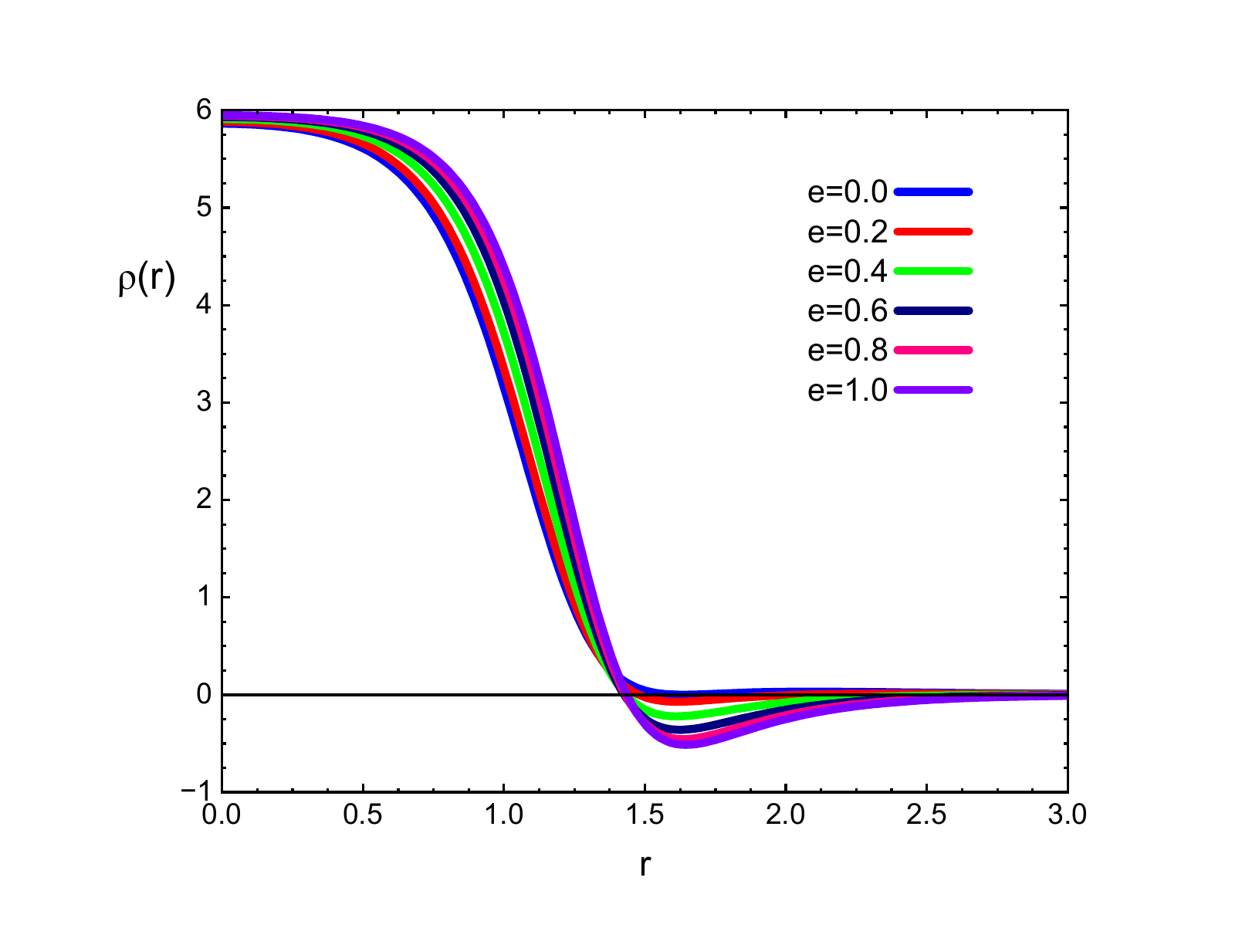}
\includegraphics[height=.285\textheight,  angle =-0]
{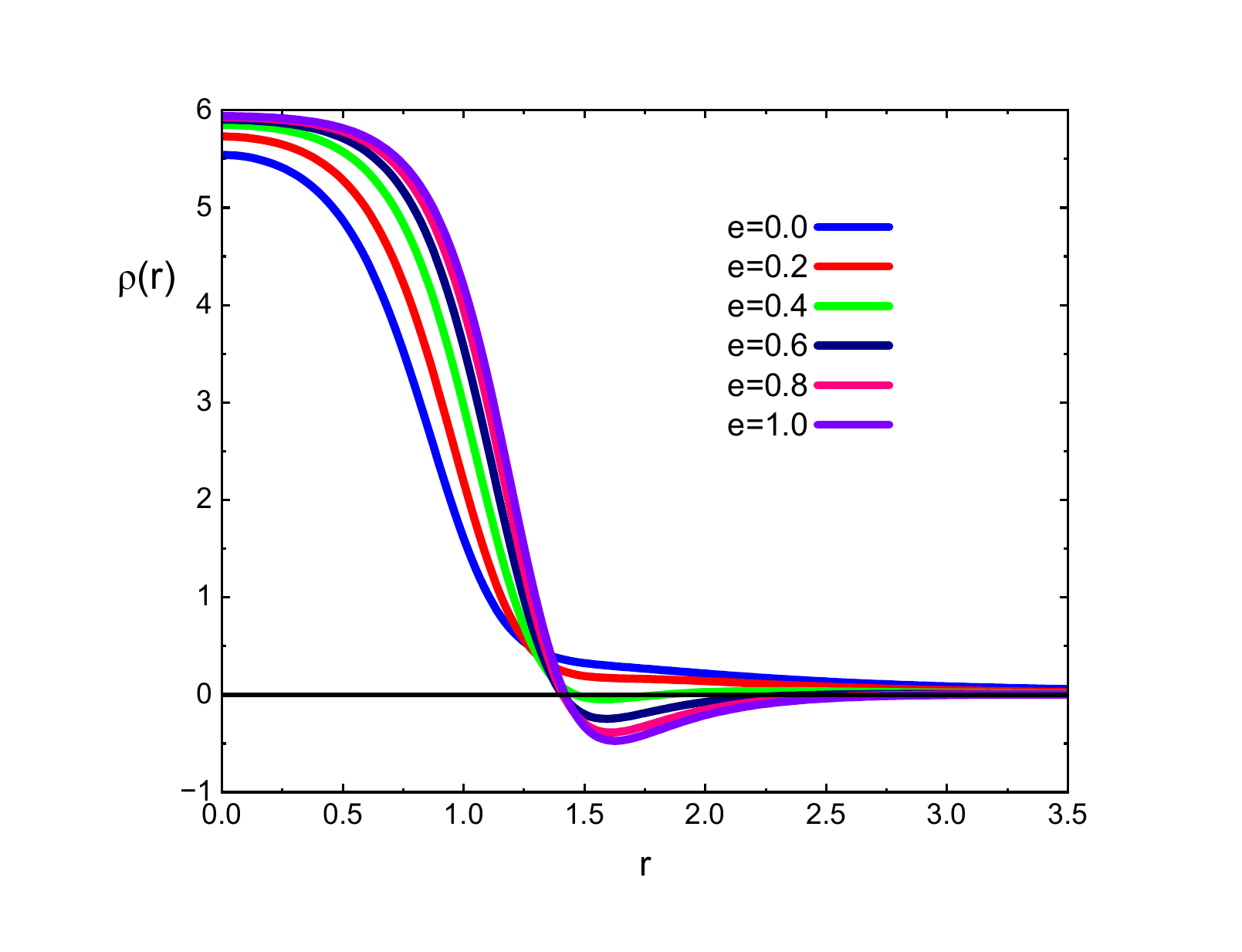}
\includegraphics[height=.285\textheight,  angle =-0]
{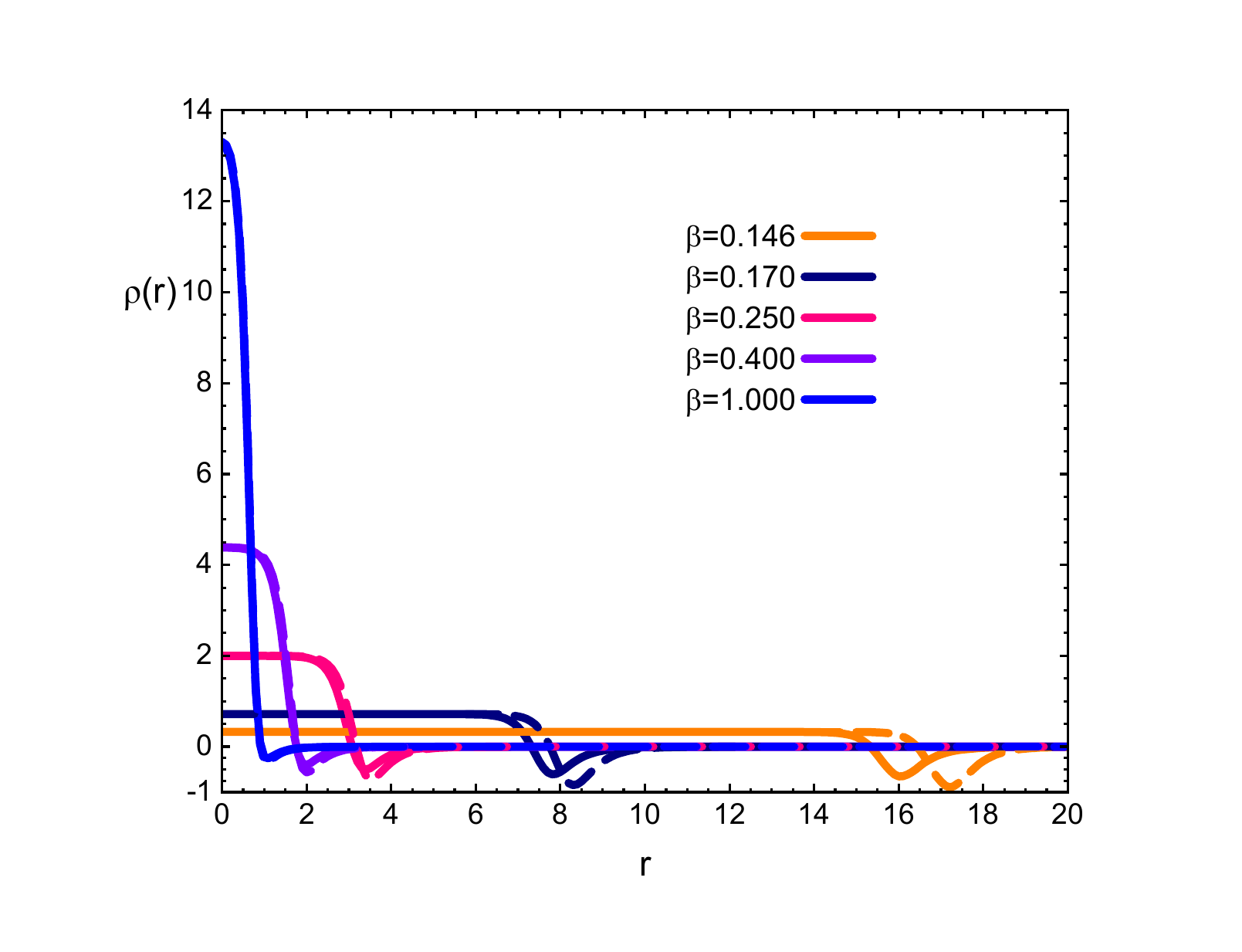}
\end{center}
\caption{\small Profiles of the function $\rho(r)$ with different values of gauge coupling $e$ at $\omega=0.8$ (upper left) and $\omega=0.995$ (upper right) for $\beta=0.5$, and for some set of values of parameter $\beta$ (bottom plot) at $e=0$ (solid line), $e=0.6$ (dashed line) for $\omega=0.5$.} 
\lbfig{fig4a}
\end{figure}
%%%%%%%%%%%%%%%%%%%%%%%%%%%%%%%%%%%%%%%%%%%%%%%%%%%%%%%%%%%%%%%%%%%%

Interestingly, the presence of domains with negative energy density and violation of the energy conditions most likely does not lead to the destabilization of solitons. 

%%%%%%%%%%%%%%%%%%%%%%%%%%%%%%%%%%%%%%%%%%%%%%%%%%%%%%%%%%%%%%%%%%%%
\begin{figure}[thb]
\begin{center}
\includegraphics[height=.285\textheight,  angle =-0]
{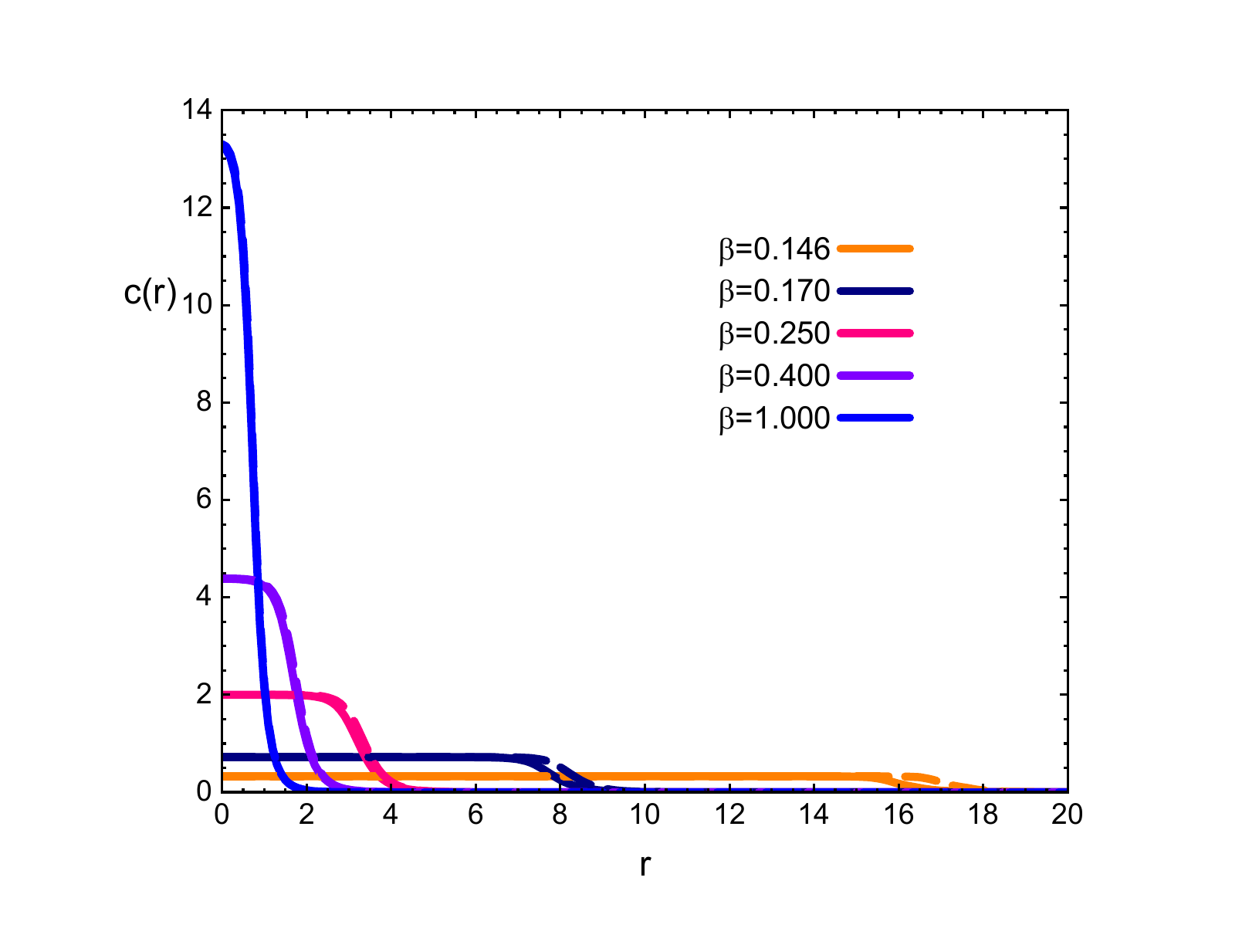}
\includegraphics[height=.285\textheight,  angle =-0]
{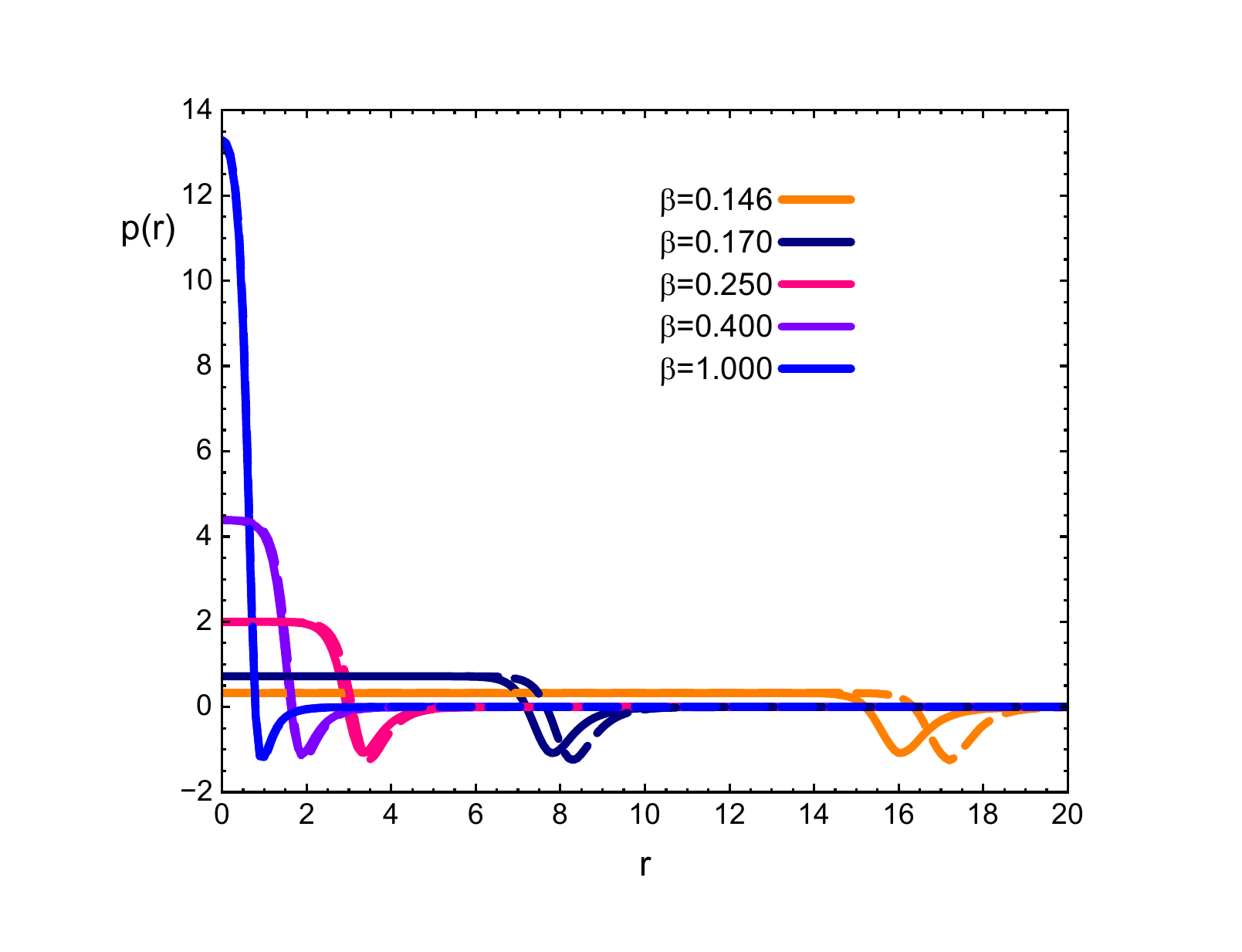}
\includegraphics[height=.285\textheight,  angle =-0]
{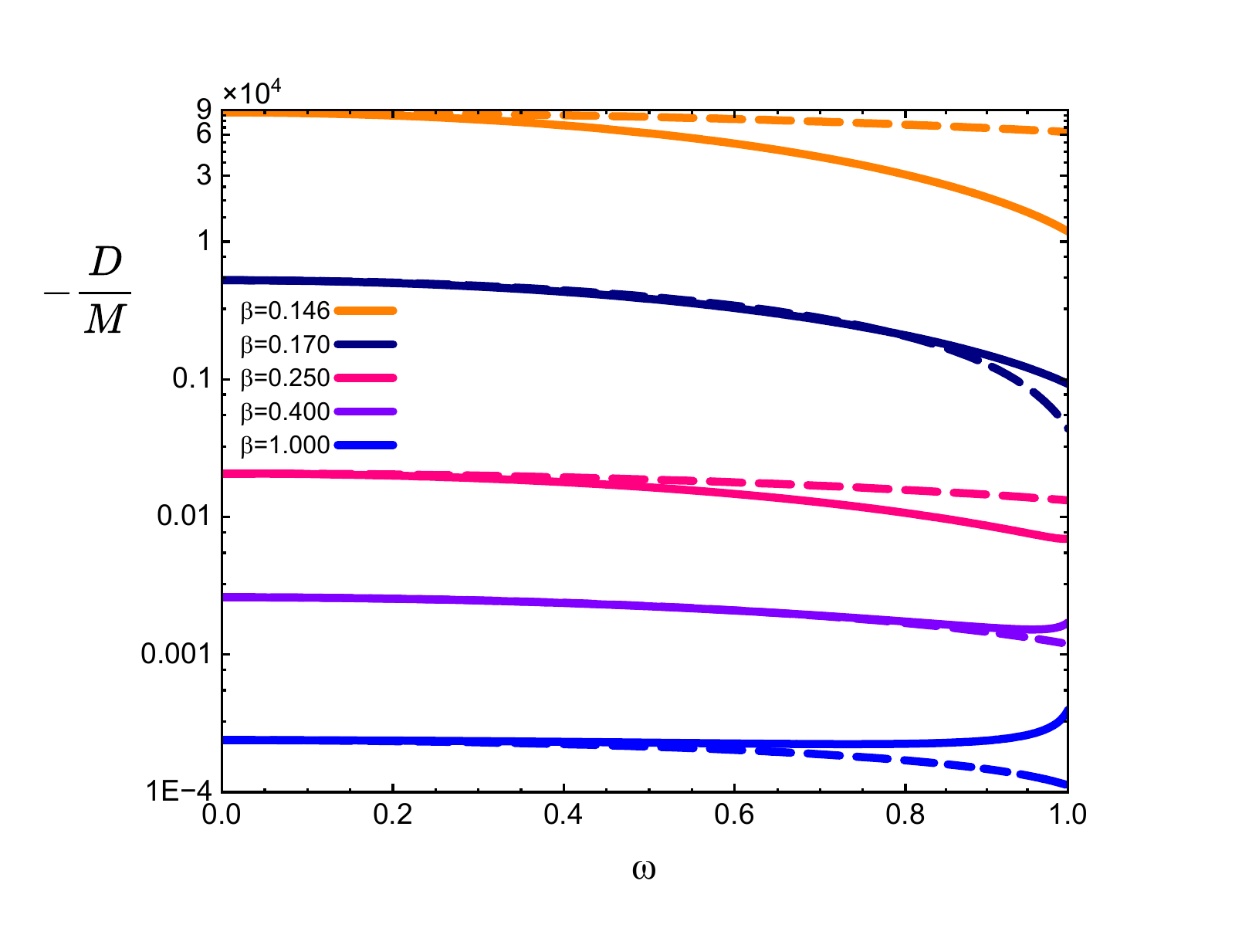}
\includegraphics[height=.285\textheight,  angle =-0]
{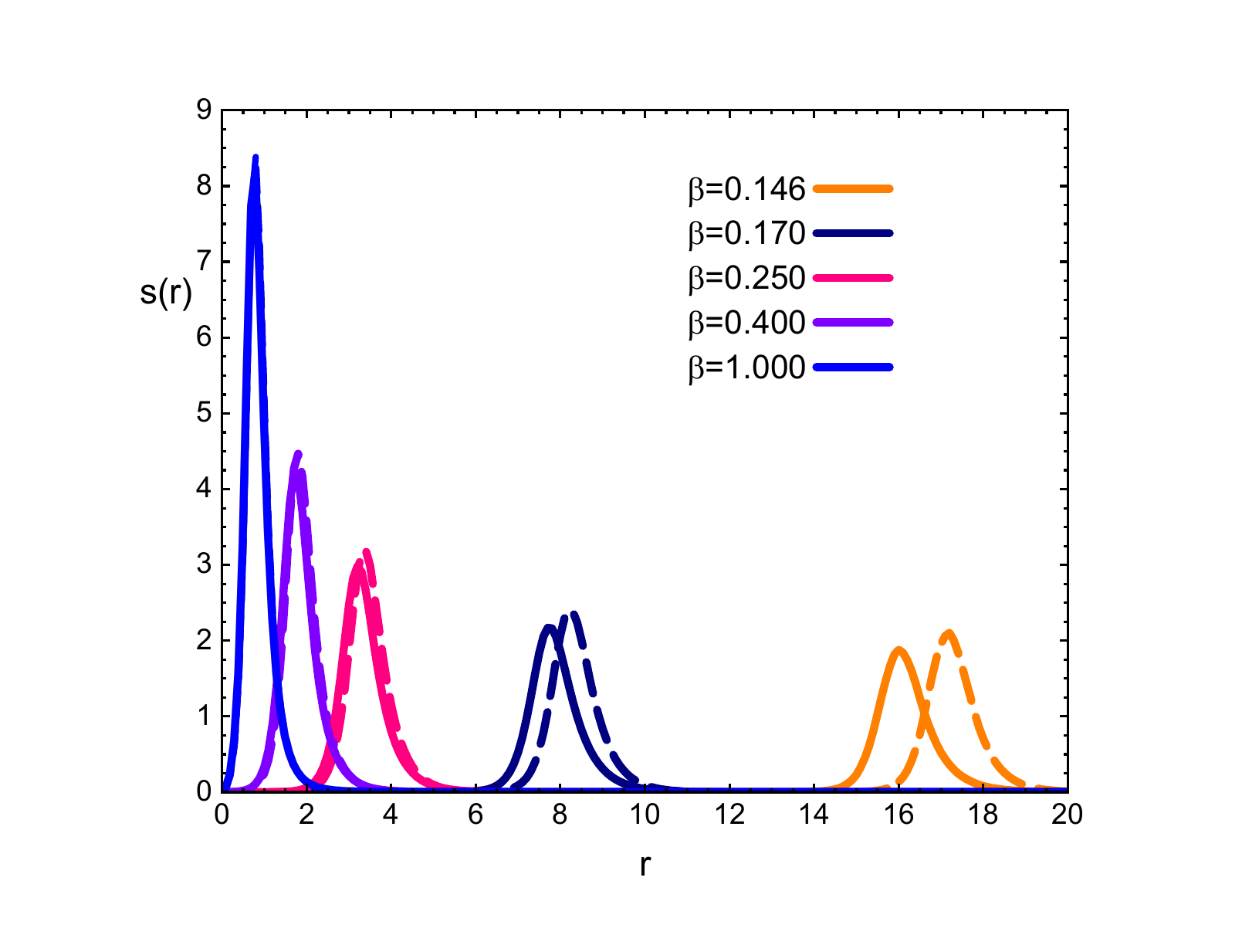}
\end{center}
\caption{\small Radial distributions of functions $c(r),p(r), s(r)$ at $\omega=0.5$ and $-\frac{D(\omega)}{M}$ for different values of the parameter $\beta$ and $e=0$ (solid lines), $e=0.2$ (dashed doted lines) and $e=0.6$ (dashed lines).} 
\lbfig{fig1b}
\end{figure}
%%%%%%%%%%%%%%%%%%%%%%%%%%%%%%%%%%%%%%%%%%%%%%%%%%%%%%%%%%%%%%%%%%%%

Indeed, in
Fig.~\ref{fig1b} we displayed the radial distributions of the pressure function $p(r)$ and the shear forces $s(r)$ \re{S_P}, and the stability criteria function $c(r)$ \re{C-criterion} for some set of values of the gauge coupling and the parameter $\beta$. We can clearly see that for all solutions the  pressure function $p(r)$ possesses no more than
one radial node while the shear force $s(r)$ is positive approaching zero both at the origin and on the spacial asymptotic. The pressure is maximal at the center of the soliton, it approaches zero from below on the spacial asymptotic.

The value of the D-term slightly varies with $\omega$, however it always remains negative, see Fig.~\ref{fig1b}, bottom left plot. 

Figs.~\ref{fig3},\ref{fig4} illustrate dependency
of the pressure function $p(r)$, the shear forces $s(r)$  and the stability criterion function $c(r)$ on the strength of the gauge coupling. For $\omega=0.8$ and $\beta=0.5$ the increase of the gauge coupling slightly increase the minimal negative value of the pressure, the central values of the functions $c(0)$ and $p(0)$ marginally increase. 
This effect becomes more noticeable as the frequency $\omega$ approaches the upper limit, see Figs.~\ref{fig4},\ref{fig6}. 
However, there are no qualitative changes in the behavior of these solutions; the pressure distribution function $p(r)$ exhibits a single zero for all parameter values, the criterion function $c(r)$ remains positive and the D-term is always negative. We may conclude that the $U(1)$ gauged soliton solutions are stress-stable. 

\begin{figure}[h!]
\begin{center}
\includegraphics[height=.285\textheight,  angle =-0]
{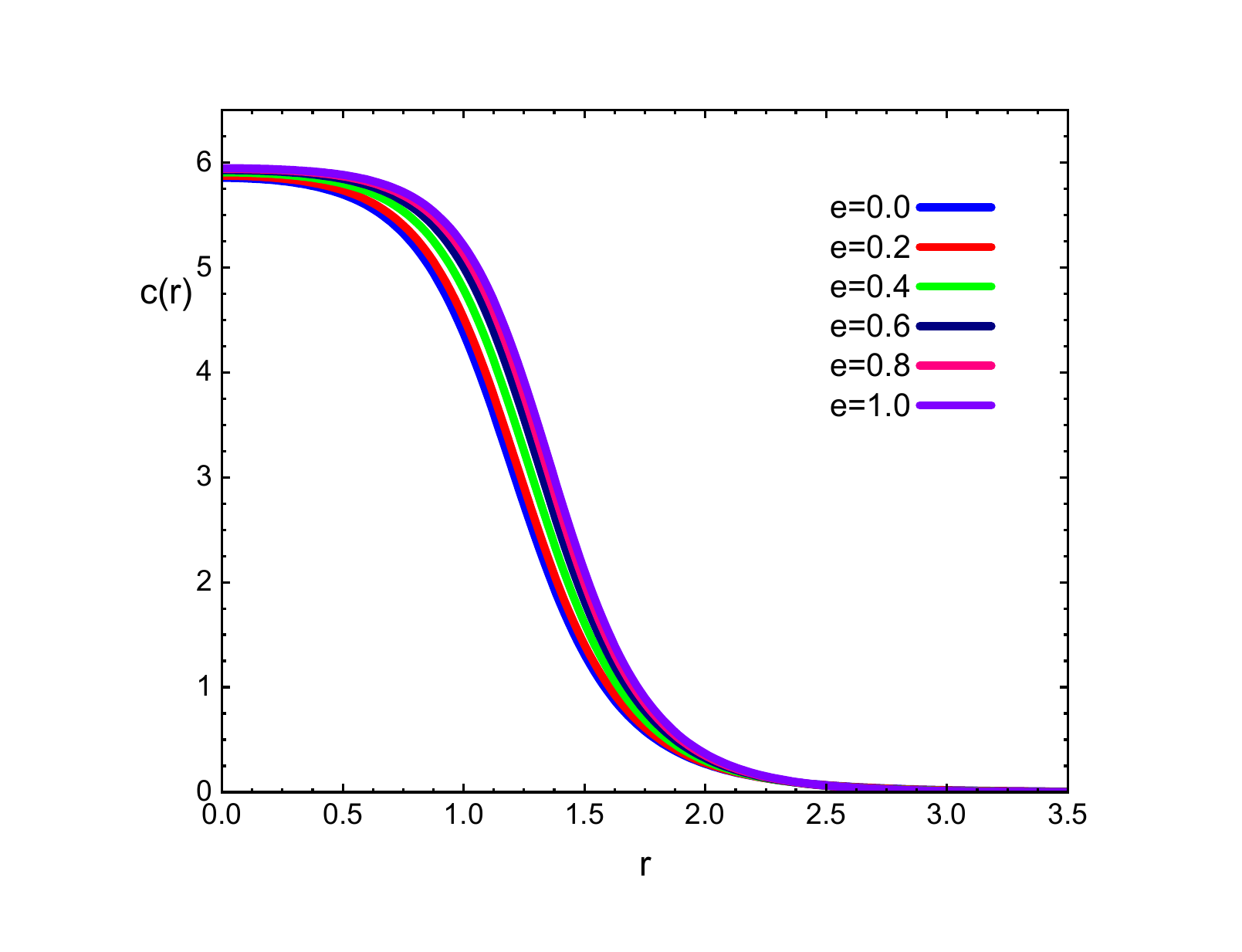}
\includegraphics[height=.285\textheight,  angle =-0]
{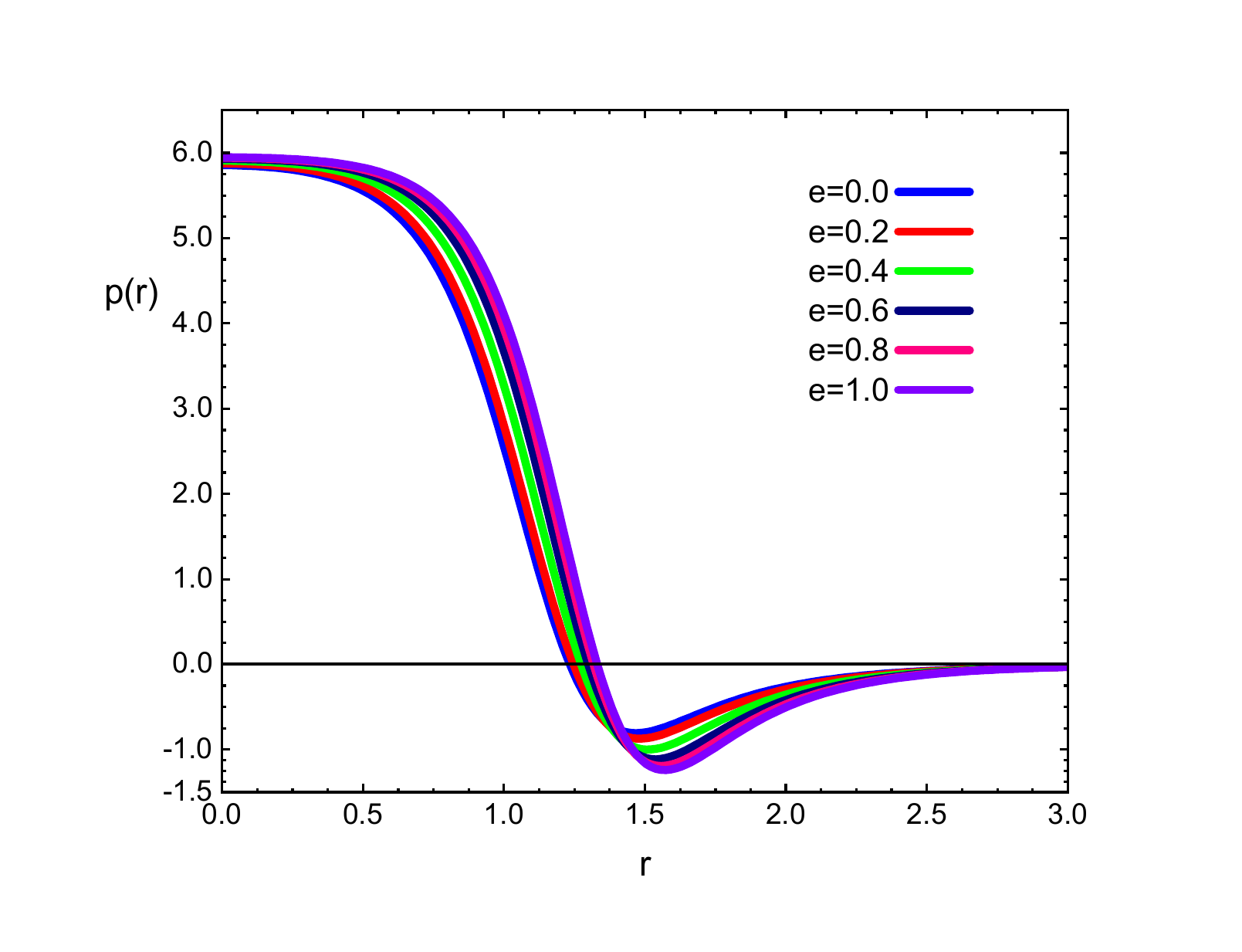}
\includegraphics[height=.285\textheight,  angle =-0]
{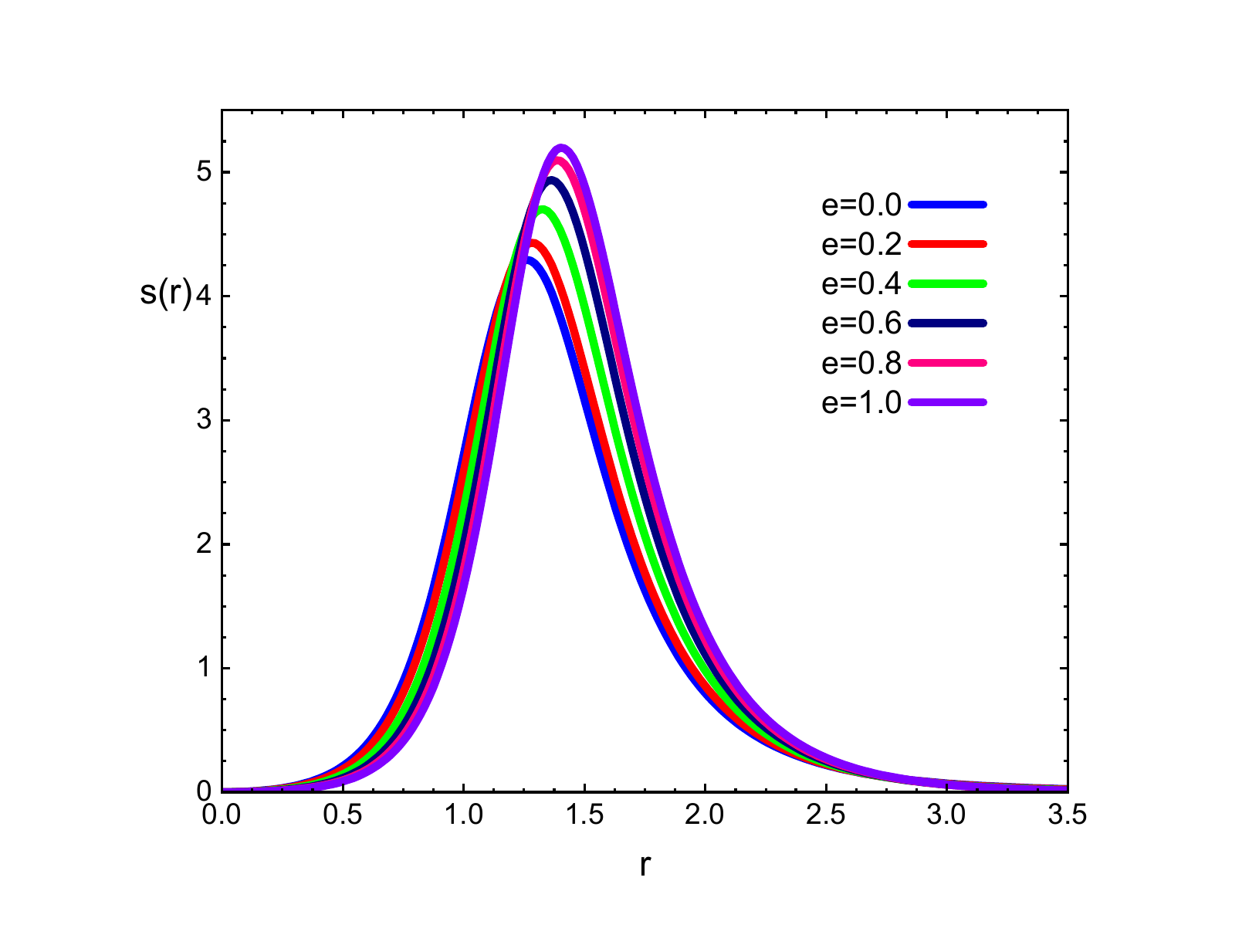}
\end{center}
\caption{\small Profiles of the functions $ c(r), p(r), s(r)$ for some set of values of the coupling $e$ at $\omega=0.8$ and $\beta=0.5$.} 
\lbfig{fig3}
\end{figure}
%%%%%%%%%%%%%%%%%%%%%%%%%%%%%%%%%%%%%%%%%%%%%%%%%%%%%%%%%%%%%%%%%%%%

It is interesting to compare mechanical properties of these solitons with those of the well established Q-balls, see \cite{Mai:2012yc,Mai:2012cx,Loiko:2022noq}. First, the radial distributions of the energy density are very different, for the usual Q-balls it is always positive with a non-zero value at the center of the configuration and a pronounced peak on the surface of the soliton. On the other hand,
qualitatively, the pressure function $p(r)$ differs little from its analogue in the case of Q-balls, it is positive inside the core of the soliton and possess at least one zero. Note that for the $U(1)$ gauged Q-balls the pressure function may possess more than one node \cite{Loiko:2022noq}, the configuration becomes unstable. 

%%%%%%%%%%%%%%%%%%%%%%%%%%%%%%%%%%%%%%%%%%%%%%%%%%%%%%%%%%%%%%%%%%%%
\begin{figure}[t!]
\begin{center}
\includegraphics[height=.285\textheight,  angle =-0]
{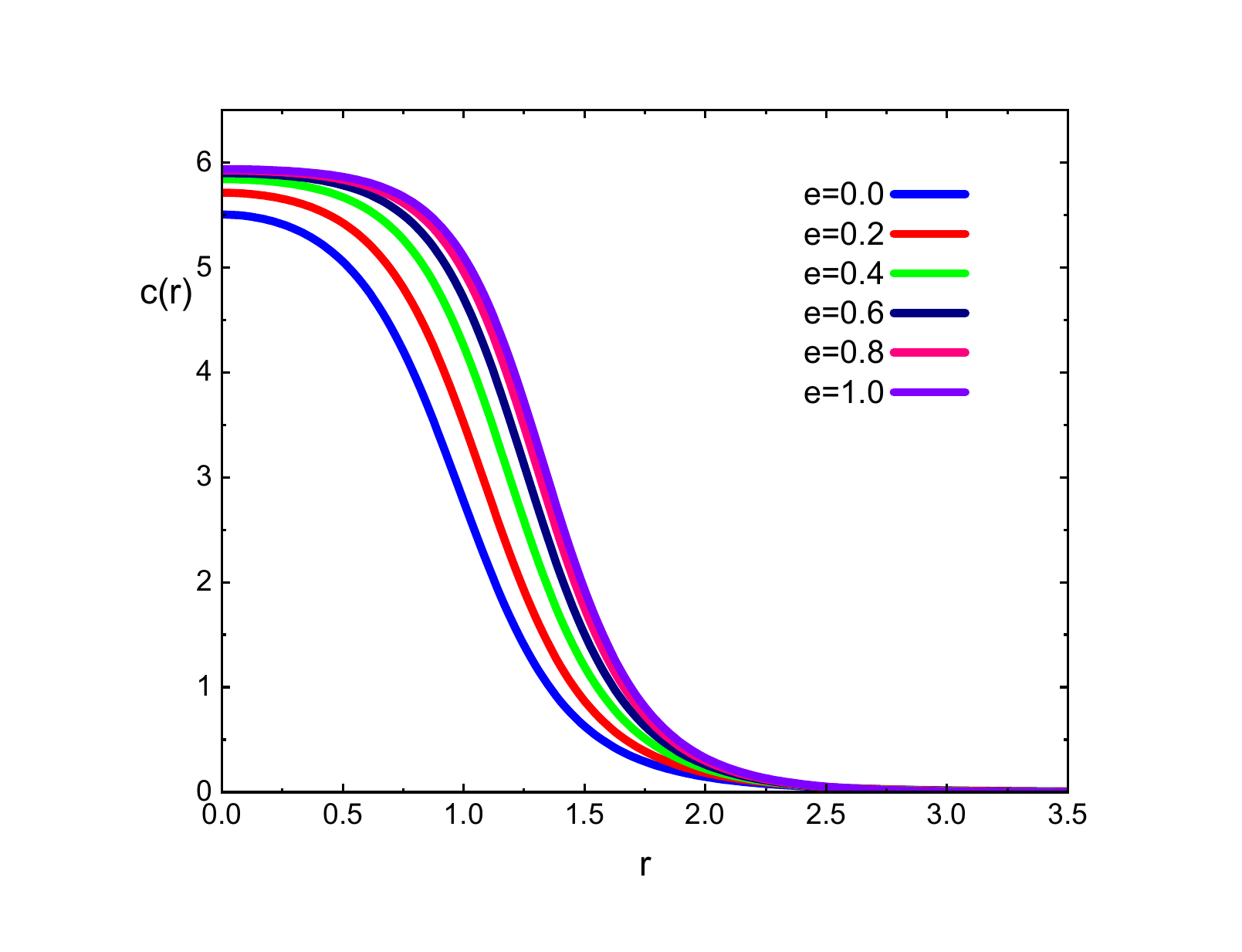}
\includegraphics[height=.285\textheight,  angle =-0]
{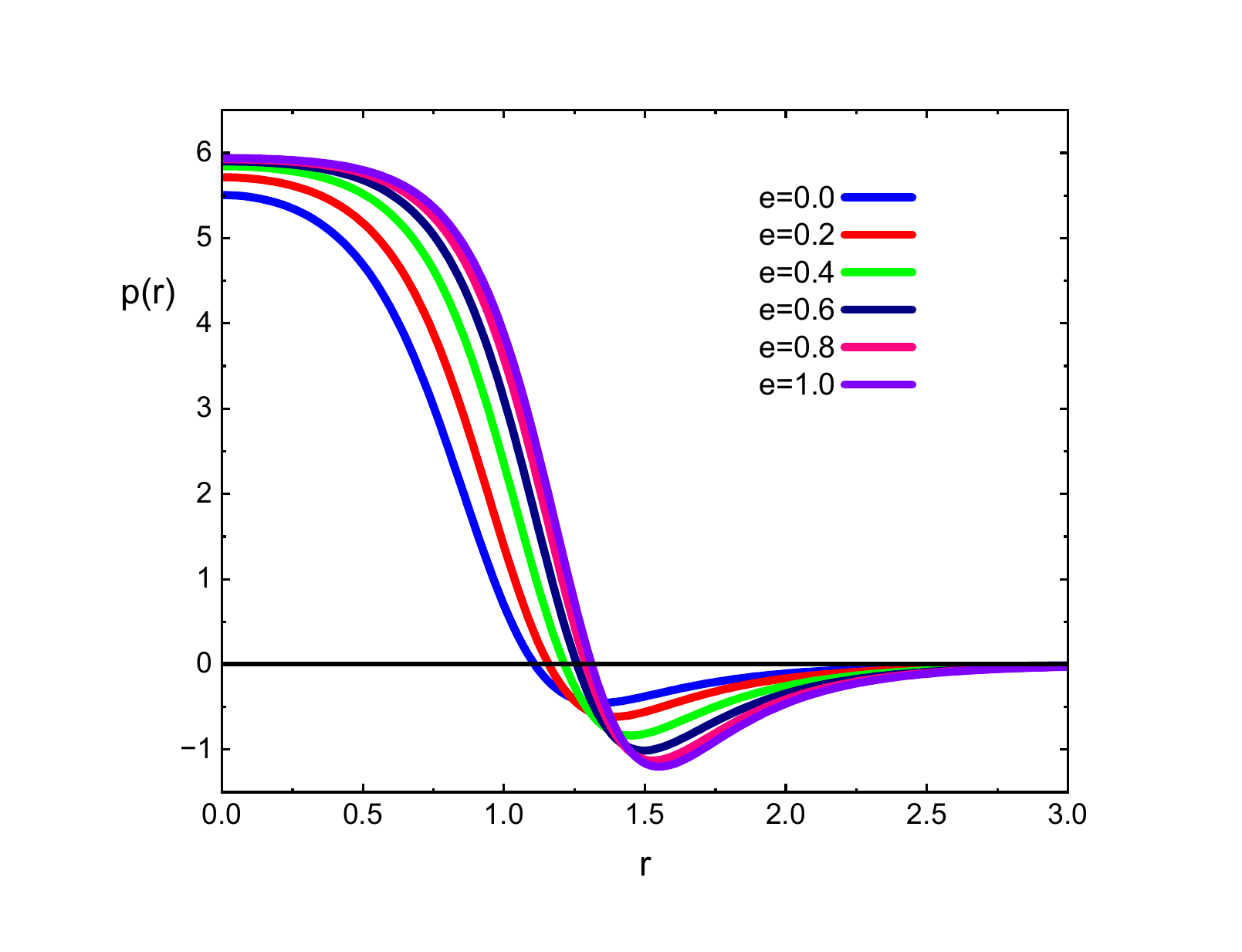}
\includegraphics[height=.285\textheight,  angle =-0]
{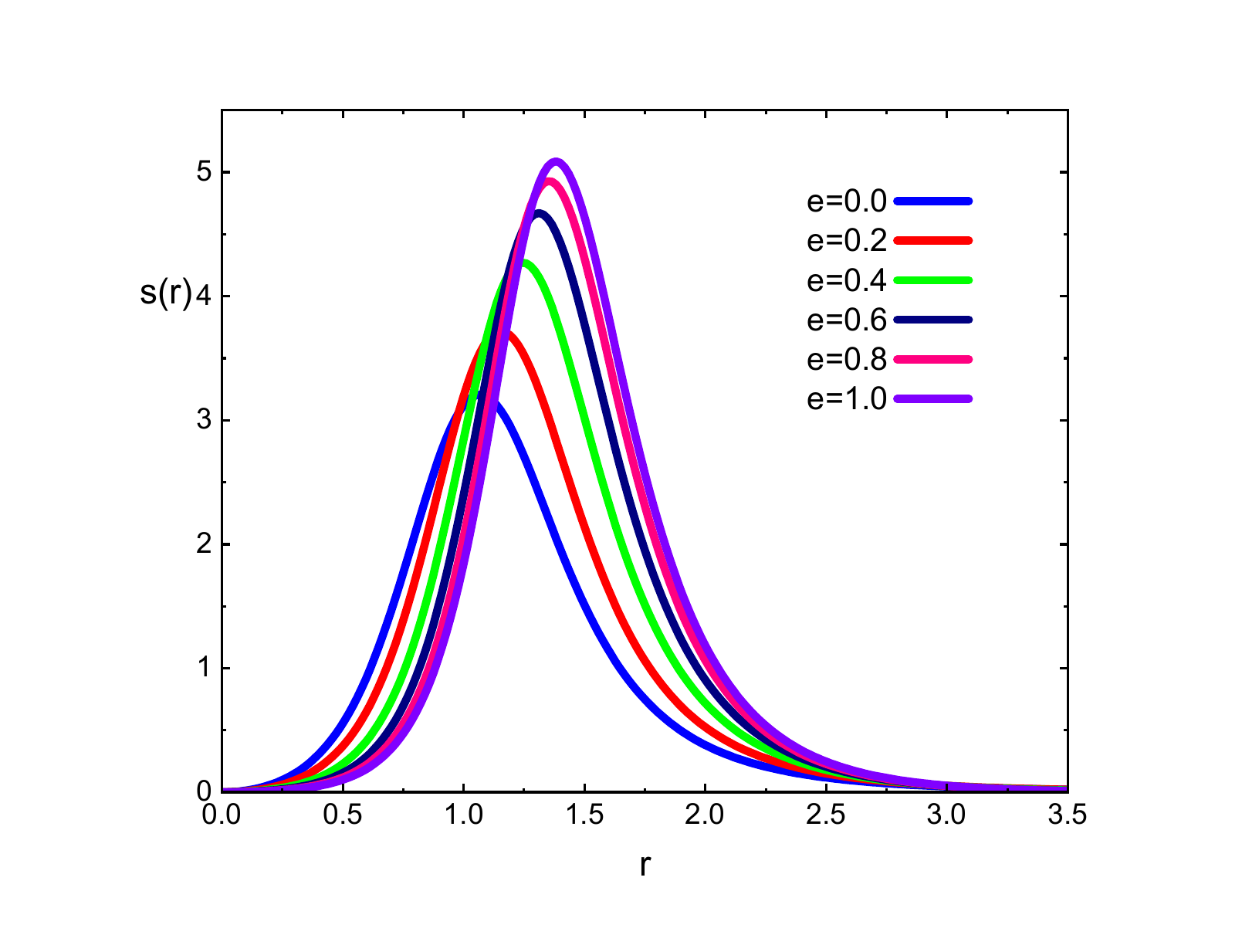}
\end{center}
\caption{\small Profiles of the functions $c(r), p(r), s(r)$ for some set of values of the coupling $e$ at $\omega=0.995$ and $\beta=0.5$.}    
\lbfig{fig4}    
\end{figure}
%%%%%%%%%%%%%%%%%%%%%%%%%
%%%%%%%%%%%%%%%%%%%%%%%%%%%%%%%%%%%%%%%

For the stable branch of $U(1)$ gauged Q-balls both the shear force function $s(r)$ and the criteria function $c(r)$ differ qualitatively little from their analogs in the non-linear $O(3)$ sigma model. However, there is an instable branch of the $U(1)$ gauged Q-balls, the corresponding shear force $s(r)$ becomes negative both inside of the core of the configuration and on the spacial
asymptotic while criteria function $c(r)$
becomes negative in these domains \cite{Loiko:2022noq}. This pattern is excluded for the $U(1)$ gauged solitons of the non-linear $O(3)$ sigma model.   

Finally, we note that  the dependence of the solutions on the parameter $\beta$ is crucial.  As $\beta$ approaches the critical value $1/8$, the stability criterion function $c(r)$, the energy density $\rho$, and the pressure distribution $p(r)$ become more spatially extended, while their central values increase. The peak of the shear force radial distribution $s(r)$ then shifts toward larger values of $r$, its amplitude decreases.

%%%%%%%%%%%%%%%%%%%%%%%%%%%%%

\begin{figure}[t!]
\begin{center}
\includegraphics[height=.285\textheight,  angle =-0]
{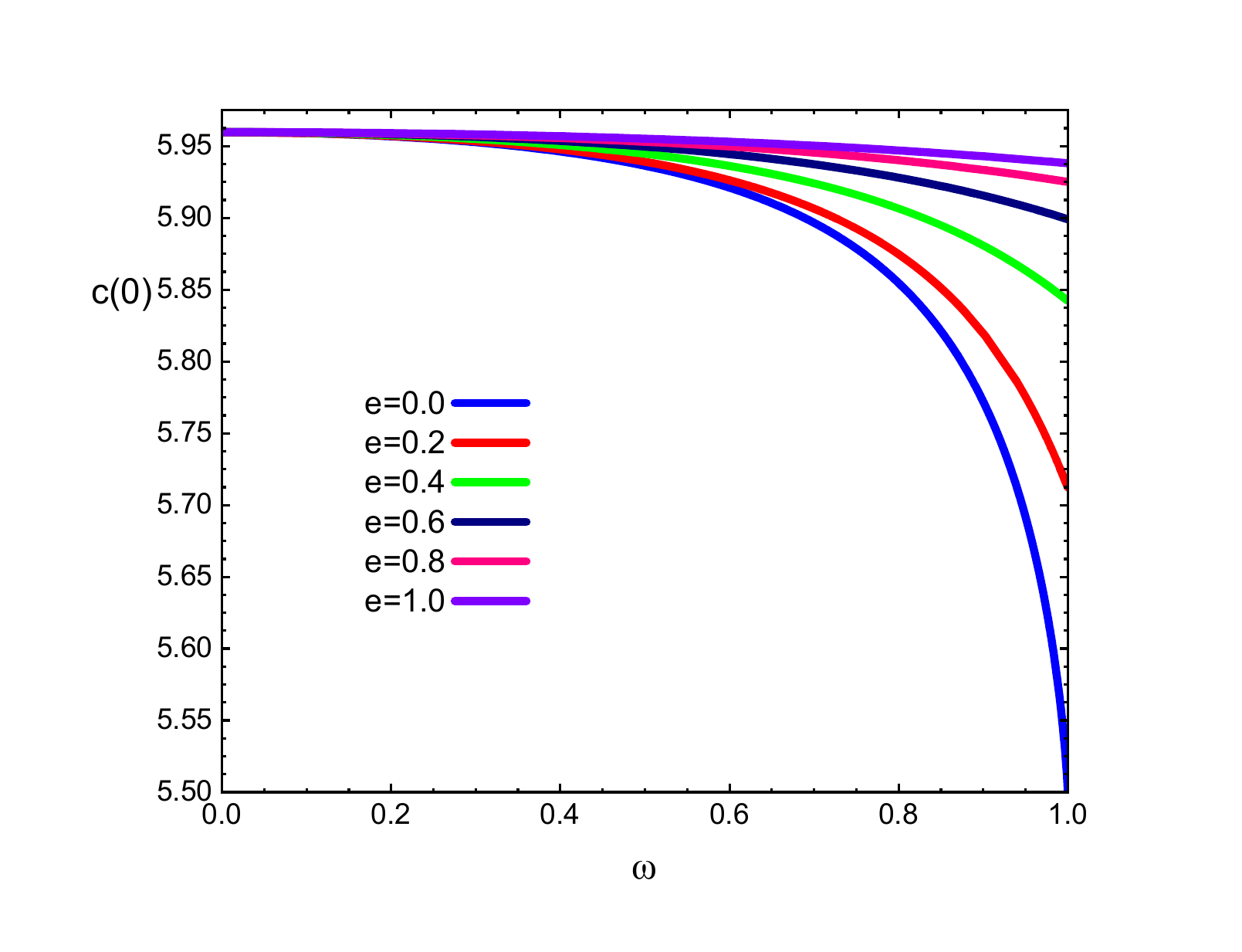}
\includegraphics[height=.285\textheight,  angle =-0]
{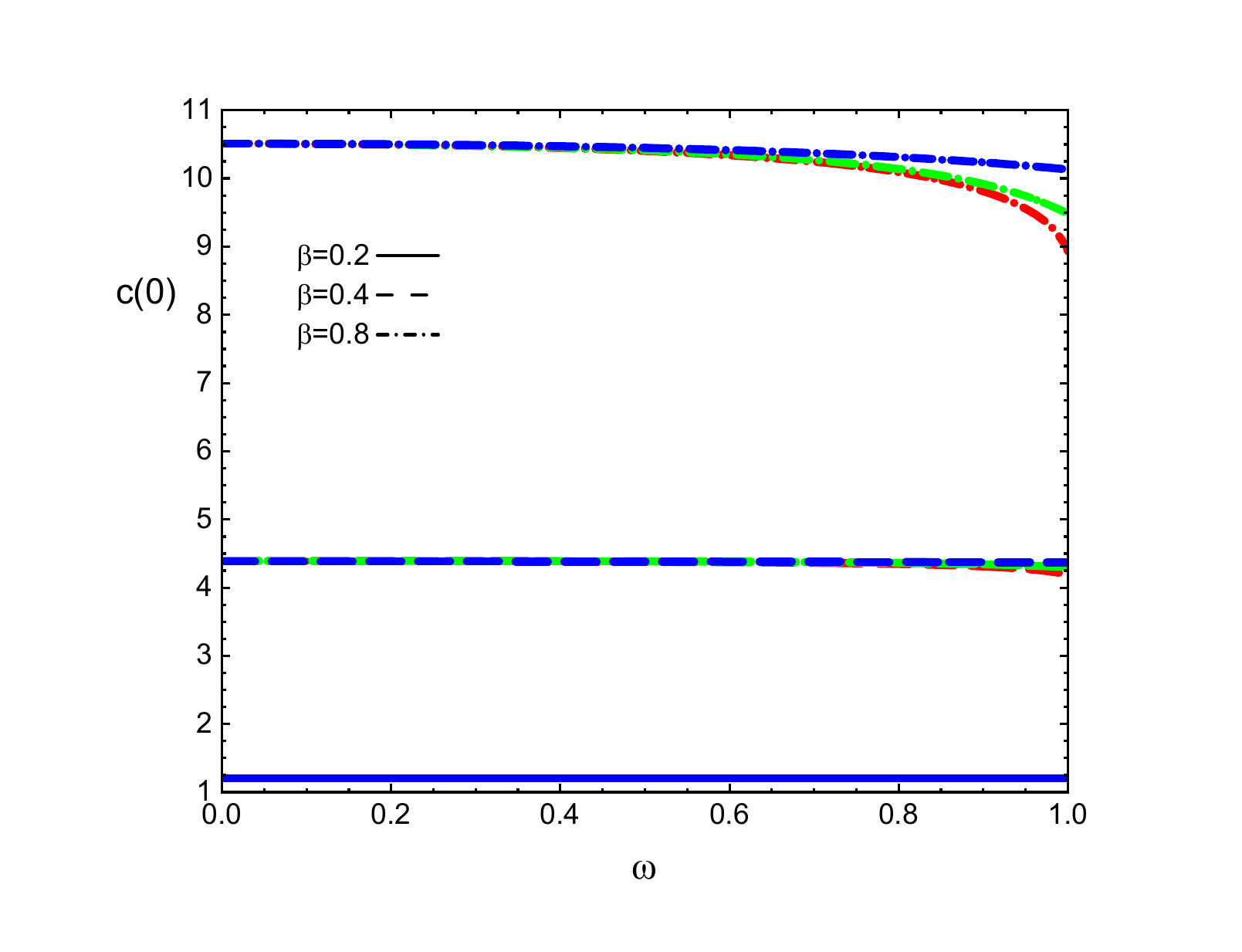}
\end{center}
\caption{\small The central value of the function $c(0)$ vs the frequency $\omega$ for some set of values of the coupling $e$ at $\beta=0.5$ (left) and for some set of values of parameter $\beta$ (right) at $e=0$ (red line), $e=0.2$ (green line), $e=0.6$ (blue line), respectively.}    
\lbfig{fig6}
\end{figure}
%%%%%%%%%%%%%%%%%%%%%%%%%%%%%%%%%%%%%%%%%%%%%%%%%%%%%%%%%%%%%%%%%%%%

%%%%%%%%%%%%%%%%%%%%%%%%%%%%%%%%%%%%%%%%%%%%%%%%%%%%%%%%%%%%%%%%%%%%

\section{Conclusions and outlook}

In this paper we have considered the problem of classical stability of spherically symmetric non-topological soliton solutions of the non-linear $O(3)$ sigma model with a symmetry-breaking potential containing a non-positive term responsible for the additional energy of the short-range interaction. Clearly, the presence of the negative energy density domain at the center of the soliton violates the weak energy condition. Further, coupling to the gauge field yields additional electrostatic repulsion which may also destabilizes the configuration.     

Our approach to the study of soliton stability is based
on a treatment of the $U(1)$ gauged non-topological soliton as an elastic medium and study of corresponding matrix elements of energy-momentum tensor, which contains
information about spatial distribution of internal forces acting in the interior of the configuration. 

Our results indicate that the von Laue stability condition is always satisfied. Further, we analyze a local stability criteria, the results of
numerical simulations demonstrate that the $U(1)$ gauged non-topological solitons of the non-linear $O(3)$ sigma  model are classically stable for the full range of values
of the parameters of the system. 

Looking ahead, several promising avenues for further
study emerge. A natural next step will be to analyze the internal mechanical properties of the axially-symmetric solitonic solutions of the $O(3)$ sigma model \cite{Ferreira:2025xey}.

An interesting open question is whether the usual Vakhitov-Kolokolov criterion of stability \cite{Vakhitov:1973lcn} can be directly applied to the 
non-topological solitons of the non-linear $O(3)$ sigma model with a harmonic time dependence.  More generally, 
there is an interesting task to study the full spectrum of linearized perturbations of these solitons. 

Finally, we note that apart from the non-topological solitons considered in this paper, the $O(3)$ sigma model in 3 spacial dimensions also supports topological solitons with a nonzero Hopf charge. 
Notably, both for Hopfion solutions and for the $O(3)$ non-topological solitons, the field is approaching the anti-vacuum $\phi^3 = -1$ at the center of configurations.   It would be interesting to investigate the possibility of the existence of topological soliton solutions in a model with a potential of the type \re{pot} and without terms with higher-order derivatives, like Skyrme term or Dzyaloshinskii–Moriya interaction.

\section*{Acknowledgments}
We thank Luiz Ferreira, Jutta Kunz and Eugen Radu for useful discussions. Y.S.
gratefully acknowledges the support from
the Hanse-Wissenschaftskolleg (Institute for Advanced Study) in Delmenhorst.
\newpage

%%%%%%%%%%%%%%%%%%%%%%%%%%%%%%
\begin{small}

\end{small}

\begin{thebibliography}{99}
%%%%%%%%%%%%%%%%%%%%%%%%%%%%%%%%%%%%%%%%%%%%%%%%%%%%%%%%%%%%%%%%%%%%%%%%%%%%%%

\bibitem{Solitons} {\it Solitons and Condensed Matter Physics},
edited by A. R. Bishop and T. Schneider (Springer-Verlag, Berlin, 1978).
%%%%%%%%%%%%%%%%%%%%%%%%%%%%%%%%%%%%%%%%%%%%%%%%%%%%%%%%%%%%%%%%%%%%%%%%%%%%%%
\bibitem{Ackerman:2017lse}P.J.~Ackerman and I.I.~Smalyukh,
%Diversity of Knot Solitons in Liquid Crystals Manifested by Linking of Preimages in Torons and Hopfions,
Phys.\ Rev.\ X , {\bf 7},  011006 (2017)
%%%%%%%%%%%%%%%%%%%%%%%%%%%%%%%%%%%%%%%%%%%%%%%%%%%%%%%%%%%%%%%%%%%%%%%%%%%%%%
\bibitem{Kosevich}A.M.~Kosevich, \textit{The crystal lattice: phonons, solitons, dislocations, superlattices},
(John Wiley $\&$ Sons, 2006).
%%%%%%%%%%%%%%%%%%%%%%%%%%%%%%%%%%%%%%%%%%%%%%%%%%%%%%%%%%%%%%%%%%%%%%%%%%%%%%
\bibitem{Mollenauer}L.F.~Mollenauer and J.P.~Gordon, {\it Solitons in Optical Fibers},
(Academic Press, 2006).
%%%%%%%%%%%%%%%%%%%%%%%%%%%%%%%%%%%%%%%%%%%%%%%%%%%%%%%%%%%%%%%%%%%%%%%%%%%%%%
\bibitem{Dauxois} T.~Dauxois and M.~Peyrard, {\it Physics of Solitons}, (Cambridge University Press, 2006)\\
M.~Peyrard,  \textit{Nonlinear Excitations in Biomolecules}, (Springer-Verlag, , 1995).
%%%%%%%%%%%%%%%%%%%%%%%%%%%%%%%%%%%%%%%%%%%%%%%%%%%%%%%%%%%%%%%%%%%%%%%%%%%%%%
%%%%%%%%%%%%%%%%%%%%%%%%%%%%%%%%%%%%%%%%%%%%%%%%%%%%%%%%%%%%%%%%%%%%%%%%%%%%%
\bibitem{Vilenkin}A.~Vilenkin and E.P.S.~Shellard,
{\it Cosmic Strings and Other Topological Defects}, (Cambridge University Press, Cambridge, England, 1994).
%%%%%%%%%%%%%%%%%%%%%%%%%%%%%%%%%%%%%%%%%%%%%%%%%%%%%%%%%%%%%%%%%%%%%%%%%%%%%%
%%%%%%%%%%%%%%%%%%%%%%%%%%%%%%%%%%%%%%%%%%%%%%%%%%%%%%%%%%%%%%%%%%%%%%%%%%%%%%
\bibitem{Manton:2004tk}
  N.~S.~Manton and P.~Sutcliffe,
  {\it 'Topological solitons',}
    Cambridge University Press, 2004.
%%%%%%%%%%%%%%%%%%%%%%%%%%%%%%%%%%%%%%%%%%%%%%%%%%%%%%%%%%%%%%%%%%%%%%%%%%%%%%
\bibitem{Shnir2018}
Y.M.~Shnir,
{\it 'Topological and Non-Topological Solitons in Scalar Field Theories'},
Cambridge University Press, 2018.
%%%%%%%%%%%%%%%%%%%%%%%%%%%%%%%%%%%%%%%%%%%%%%%%%%%%%%%%%%%%%%%%%%%%%%%%%%%%%%
%\cite{Gell-Mann:1960mvl}
\bibitem{Gell-Mann:1960mvl}
M.~Gell-Mann and M.~Levy,
%``The axial vector current in beta decay,''
Nuovo Cim. \textbf{16} (1960), 705
%%%%%%%%%%%%%%%%%%%%%%%%%%%%%%%%%%%%%%%%%%%%%%%%%%%%%%%%%%%%%%%%%%%%%%%%%%%%%%
%%%%%%%%%%%%%%%%%%%%%%%%%%%%%%%%%%%%%%%%%%%%%%%%%%%%%%%%%%%%%%%%%%%%%%%%%%%%%%
%\cite{Polyakov:1975yp}
\bibitem{Polyakov:1975yp}
A.~M.~Polyakov and A.~A.~Belavin,
%``Metastable States of Two-Dimensional Isotropic Ferromagnets,''
JETP Lett. \textbf{22} (1975), 245-248
%724 citations counted in INSPIRE as of 15 Jan 2026
%%%%%%%%%%%%%%%%%%%%%%%%%%%%%%%%%%%%%%%%%%%%%%%%%%%%%%%%%%%%%%%%%%%%%%%%%%%%%%
%%%%%%%%%%%%%%%%%%%%%%%%%%%%%%%%%%%%%%%%%%%%%%%%%%%%%%%%%%%%%%%%%%%%%%%%%%%%%%
%\cite{Derrick:1964ww}
\bibitem{Derrick:1964ww}
G.~H.~Derrick,
%``Comments on nonlinear wave equations as models for elementary particles,''
J. Math. Phys. \textbf{5} (1964), 1252-1254
%%%%%%%%%%%%%%%%%%%%%%%%%%%%%%%%%%%%%%%%%%%%%%%%%%%%%%%%%%%%%%%%%%%%%%%%%%%%%%
%\cite{Faddeev:1976pg}
\bibitem{Faddeev:1976pg}
L.~D.~Faddeev,
%``Some Comments on the Many Dimensional Solitons,''
Lett. Math. Phys. \textbf{1} (1976), 289
%%%%%%%%%%%%%%%%%%%%%%%%%%%%%%%%%%%%%%%%%%%%%%%%%%%%%
%\cite{Gladikowski:1996mb}
\bibitem{Gladikowski:1996mb}
J.~Gladikowski and M.~Hellmund,
%``Static solitons with nonzero Hopf number,''
Phys. Rev. D \textbf{56} (1997), 5194-5199
%%%%%%%%%%%%%%%%%%%%%%%%%%%%%%%%%%%%%%%%%%%%%%%%%%%%%
%\cite{Battye:1998pe}
\bibitem{Battye:1998pe}
R.~A.~Battye and P.~M.~Sutcliffe,
%``Knots as stable soliton solutions in a three-dimensional classical field theory.,''
Phys. Rev. Lett. \textbf{81} (1998), 4798-4801
%%%%%%%%%%%%%%%%%%%%%%%%%%%%%%%%%%%%%%%%%%%%%%%%%%%%%
%\cite{Harland:2013uk}
\bibitem{Harland:2013uk}
D.~Harland, J.~J{\"a}ykk{\"a}, Y.~Shnir and M.~Speight,
%``Isospinning hopfions,''
J. Phys. A \textbf{46} (2013), 225402
%%%%%%%%%%%%%%%%%%%%%%%%%%%%%%%%%%%%%%%%%%%%%%%%%%%%%
%\cite{Battye:2013xf}
\bibitem{Battye:2013xf}
R.~A.~Battye and M.~Haberichter,
%``Classically Isospinning Hopf Solitons,''
Phys. Rev. D \textbf{87} (2013) no.10, 105003
%%%%%%%%%%%%%%%%%%%%%%%%%%%%%%%%%%%%%%%%%%%%%%%%%%%%%
%\cite{Samoilenka:2017hmn}
\bibitem{Samoilenka:2017hmn}
A.~Samoilenka and Y.~Shnir,
%``Fractional Hopfions in the Faddeev-Skyrme model with a symmetry breaking potential,''
JHEP \textbf{09} (2017), 029
%%%%%%%%%%%%%%%%%%%%%%%%%%%%%%%%%%%%%%%%%%%%%%%%%%%%%
%%%%%%%%%%%%%%%%%%%%%%%%%%%%%%%%%%%%%%%%%%%%%%%%%%%%%%%%%%%%%%%%%%%%%%%%%%%%%%
%\cite{Abraham:1991ki}
\bibitem{Abraham:1991ki}
E.~Abraham,
%``Nonlinear sigma models and their Q lump solutions,''
Phys. Lett. B \textbf{278} (1992), 291-296

%%%%%%%%%%%%%%%%%%%%%%%%%%%%%%%%%%%%%%%%%%%%%%%%%%%%%%%%%%%%%%%%%%%%%%%%%%%%%%
%\cite{Ward:2003un}
\bibitem{Ward:2003un}
R.~S.~Ward,
%``Topological Q solitons,''
J. Math. Phys. \textbf{44} (2003), 3555-3561

%%%%%%%%%%%%%%%%%%%%%%%%%%%%%%%%%%%%%%%%%%%%%%%%%%%%%%%%%%%%%%%%%%%%%%%%%%%%%%
%\cite{Amari:2024pnw}
\bibitem{Amari:2024pnw}
Y.~Amari, S.~Antsipovich, M.~Nitta and Y.~Shnir,
%``Isospinning CP2 solitons,''
Phys. Rev. D \textbf{110} (2024) no.8, 085008
%%%%%%%%%%%%%%%%%%%%%%%%%%%%%%%%%%%%%%%%%%%%%%%%%%%%%%%%%%%%%%%%%%%%%%%%%%%%%%
%\cite{Antsipovich:2025liy}
\bibitem{Antsipovich:2025liy}
S.~Antsipovich,
%``Hamiltonian approach to isospinning {\ensuremath{\mathbb{C}}}P2 solitons,''
Int. J. Mod. Phys. A \textbf{40} (2025) no.31, 2550147
%%%%%%%%%%%%%%%%%%%%%%%%%%%%%%%%%%%%%%%%%%%%%%%%%%%%%%%%%%%%%%%%%%%%%%%%%%%%%%
\bibitem{Radu:2008pp}E.~Radu and M.S.~Volkov,
%Existence of stationary, non-radiating ring solitons in field theory: knots and vortons,
Phys.\ Rept.\  {\bf 468}  (2008)  101.
%%%%%%%%%%%%%%%%%%%%%%%%%%%%%%%%%%%%%%%%%%%%%%%%%%%%%%%%%%%%%%%%%%%%%%%%%%%%%%
%%%%%%%%%%%%%%%%%%%%%%%%%%%%%%%%%%%%%%%%%%%%%%%%%%%%%%%%%%%%%%%%%%%%%%%%%%%%%%
\bibitem{Rosen}G.~Rosen, J.\ Math.\ Phys.\ {\bf 9} (1968) 996, 999
%%%%%%%%%%%%%%%%%%%%%%%%%%%%%%%%%%%%%%%%%%%%%%%%%%%%%%%%%%%%%%%%%%%%%%%%%%%%%%
\bibitem{Friedberg:1976me}
  R.~Friedberg, T.~D.~Lee and A.~Sirlin,
  %``A Class of Scalar-Field Soliton Solutions in Three Space Dimensions,''
  Phys.\ Rev.\ D {\bf 13} (1976) 2739
%%%%%%%%%%%%%%%%%%%%%%%%%%%%%%%%%%%%%%%%%%%%%%%%%%%%%%%%%%%%%%%%%%%%%%%%%%%%%%
\bibitem{Coleman:1985ki}
  S.~R.~Coleman,
  %``Q Balls,''
  Nucl.\ Phys.\ B {\bf 262} (1985) 263
   Erratum: [Nucl.\ Phys.\ B {\bf 269} (1986) 744].
%%%%%%%%%%%%%%%%%%%%%%%%%%%%%%%%%%%%%%%%%%%%%%%%%%%%%%%%%%%%%%%%%%%%%%%%%%%%%%
\bibitem{Lee:1991ax}
  T.~D.~Lee and Y.~Pang,
  %``Nontopological solitons,''
  Phys.\ Rept.\  {\bf 221} (1992) 251
%%%%%%%%%%%%%%%%%%%%%%%%%%%%%%%%%%%%%%%%%%%%%%%%%%%%%%%%%%%%%%%%%%%%%%%%%%%%%%
%%%%%%%%%%%%%%%%%%%%%%%%%%%%%%%%%%%%%%%%%%%%%%%%%%%%%%%%%%%%%%%%%%%%%%%%%%%%%%
%\cite{Halavanau:2013vsa}
\bibitem{Halavanau:2013vsa}
A.~Halavanau and Y.~Shnir,
%``Isorotating Baby Skyrmions,''
Phys. Rev. D \textbf{88} (2013) no.8, 085028
%%%%%%%%%%%%%%%%%%%%%%%%%%%%%%%%%%%%%%%%%%%%%%%%%%%%%%%%%%%%%%%%%%%%%%%%%%%%%%
%\cite{Battye:2013tka}
\bibitem{Battye:2013tka}
R.~A.~Battye and M.~Haberichter,
%``Isospinning baby Skyrmion solutions,''
Phys. Rev. D \textbf{88} (2013), 125016
%%%%%%%%%%%%%%%%%%%%%%%%%%%%%%%%%%%%%%%%%%%%%%%%%%%%%%%%%%%%%%%%%%%%%%%%%%%%%%
%\cite{Verbin:2007fa}
\bibitem{Verbin:2007fa}
Y.~Verbin,
%``Sigma model Q-balls and Q-stars,''
Phys. Rev. D \textbf{76} (2007), 085018
%%%%%%%%%%%%%%%%%%%%%%%%%%%%%%%%%%%%%%%%%%%%%%%%%%%%%
%\cite{Ferreira:2025xey}
\bibitem{Ferreira:2025xey}
L.~A.~Ferreira, A.~Mikhaliuk and Y.~Shnir,
%``U(1) gauged nontopological solitons in the 3+1 dimensional O(3) sigma-model,''
Phys. Rev. D \textbf{112} (2025) no.2, 025003
%%%%%%%%%%%%%%%%%%%%%%%%%%%%%%%%%%%%%%%%%%%%%%%%%%%%%
%\cite{Gladikowski:1995sc}
\bibitem{Gladikowski:1995sc}
J.~Gladikowski, B.~M.~A.~G.~Piette and B.~J.~Schroers,
%``Skyrme-Maxwell solitons in (2+1)-dimensions,''
Phys. Rev. D \textbf{53} (1996), 844-851
%%%%%%%%%%%%%%%%%%%%%%%%%%%%%%%%%%%%%%%%%%%%%%%%%%%%%%%%%%%%%%%%%%%%%%%%%%%%%%
%\cite{Cavalcante:1996zg}
\bibitem{Cavalcante:1996zg}
F.~S.~A.~Cavalcante, M.~S.~Cunha and C.~A.~S.~Almeida,
%``Vortices in a nonminimal Maxwell Chern-Simons O(3) sigma model,''
Phys. Lett. B \textbf{475} (2000), 315-323
%%%%%%%%%%%%%%%%%%%%%%%%%%%%%%%%%%%%%%%%%%%%%%%%%%%%%%%%%%%%%%%%%%%%%%%%%%%%%%
%\cite{Bogolubskaya:1989ha}
\bibitem{Bogolubskaya:1989ha}
A.~A.~Bogolubskaya and I.~L.~Bogolubsky,
%``Stationary Topological Solitons in the Two-dimensional Anisotropic Heisenberg Model With a Skyrme Term,''
Phys. Lett. A \textbf{136} (1989), 485-488

%%%%%%%%%%%%%%%%%%%%%%%%%%%%%%%%%%%%%%%%%%%%%%%%%%%%%%%%%%%%%%%%%%%%%%%%%%%%%%
%\cite{Bogolyubskaya:1989fz}
\bibitem{Bogolyubskaya:1989fz}
A.~A.~Bogolyubskaya and I.~L.~Bogolyubsky,
%``ON STATIONARY TOPOLOGICAL SOLITONS IN TWO-DIMENSIONAL ANISOTROPIC HEISENBERG MODEL,''
Lett. Math. Phys. \textbf{19} (1990), 171-177

%%%%%%%%%%%%%%%%%%%%%%%%%%%%%%%%%%%%%%%%%%%%%%%%%%%%%%%%%%%%%%%%%%%%%%%%%%%%%%
%\cite{Piette:1994jt}
\bibitem{Piette:1994jt}
B.~M.~A.~G.~Piette, W.~J.~Zakrzewski, H.~J.~W.~Mueller-Kirsten and D.~H.~Tchrakian,
%``A Modified Mottola-Wipf model with sphaleron and instanton fields,''
Phys. Lett. B \textbf{320} (1994), 294-298

%%%%%%%%%%%%%%%%%%%%%%%%%%%%%%%%%%%%%%%%%%%%%%%%%%%%%%%%%%%%%%%%%%%%%%%%%%%%%%
%\cite{Piette:1994ug}
\bibitem{Piette:1994ug}
B.~M.~A.~G.~Piette, B.~J.~Schroers and W.~J.~Zakrzewski,
%``Multi - solitons in a two-dimensional Skyrme model,''
Z. Phys. C \textbf{65} (1995), 165-174

%%%%%%%%%%%%%%%%%%%%%%%%%%%%%%%%%%%%%%%%%%%%%%%%%%%%%%%%%%%%%%%%%%%%%%%%%%%%%%
%\cite{Samoilenka:2018oil}
\bibitem{Samoilenka:2018oil}
A.~Samoilenka and Y.~Shnir,
%``Magnetic Hopfions in the Faddeev-Skyrme-Maxwell model,''
Phys. Rev. D \textbf{97} (2018) no.12, 125014
%%%%%%%%%%%%%%%%%%%%%%%%%%%%%%%%%%%%%%%%%%%%%%%%%%%%%%%%%%%%%%%%%%%%%%%%%%%%%%
%\cite{Samoilenka:2017fwj}
\bibitem{Samoilenka:2017fwj}
A.~Samoilenka and Y.~Shnir,
%``Gauged merons,''
Phys. Rev. D \textbf{97} (2018) no.4, 045004
%%%%%%%%%%%%%%%%%%%%%%%%%%%%%%%%%%%%%%%%%%%%%%%%%%%%%%%%%%%%%%%%%%%%%%%%%%%%%%
%\cite{Samoilenka:2016wys}
\bibitem{Samoilenka:2016wys}
A.~Samoilenka and Y.~Shnir,
%``Gauged baby Skyrme model with a Chern-Simons term,''
Phys. Rev. D \textbf{95} (2017) no.4, 045002
%%%%%%%%%%%%%%%%%%%%%%%%%%%%%%%%%%%%%%%%%%%%%%%%%%%%%%%%%%%%%%%%%%%%%%%%%%%%%%
%\cite{Samoilenka:2015bsf}
\bibitem{Samoilenka:2015bsf}
A.~Samoilenka and Y.~Shnir,
%``Gauged multisoliton baby Skyrme model,''
Phys. Rev. D \textbf{93} (2016) no.6, 065018
%%%%%%%%%%%%%%%%%%%%%%%%%%%%%%%%%%%%%%%%%%%%%%%%%%%%%%%%%%%%%%%%%%%%%%%%%%%%%%
%\cite{Shnir:2014mfa}
\bibitem{Shnir:2014mfa}
Y.~Shnir and G.~Zhilin,
%``Gauged Hopfions,''
Phys. Rev. D \textbf{89} (2014) no.10, 105010
%%%%%%%%%%%%%%%%%%%%%%%%%%%%%%%%%%%%%%%%%%%%%%%%%%%%%%%%%%%%%%%%%%%%%%%%%%%%%%
%\cite{Polyakov:2002yz}
\bibitem{Polyakov:2002yz}
M.~V.~Polyakov,
%``Generalized parton distributions and strong forces inside nucleons and nuclei,''
Phys. Lett. B \textbf{555} (2003), 57-62
%%%%%%%%%%%%%%%%%%%%%%%%%%%%%%%%%%%%%%%%%%%%%%%%%%%%%%%%%%%%%%%%%%%%%%%%%%%%%%
%\cite{Polyakov:2018zvc}
\bibitem{Polyakov:2018zvc}
M.~V.~Polyakov and P.~Schweitzer,
%``Forces inside hadrons: pressure, surface tension, mechanical radius, and all that,''
Int. J. Mod. Phys. A \textbf{33} (2018) no.26, 1830025
%%%%%%%%%%%%%%%%%%%%%%%%%%%%%%%%%%%%%%%%%%%%%%%%%%%%%%%%%%%%%%%%%%%%%%%%%%%%%%
%\cite{Mai:2012yc}
\bibitem{Mai:2012yc}
M.~Mai and P.~Schweitzer,
%``Energy momentum tensor, stability, and the D-term of Q-balls,''
Phys. Rev. D \textbf{86} (2012), 076001
%%%%%%%%%%%%%%%%%%%%%%%%%%%%%%%%%%%%%%%%%%%%%%%%%%%%%%%%%%%%%%%%%%%%%%%%%%%%%%
%\cite{Mai:2012cx}
\bibitem{Mai:2012cx}
M.~Mai and P.~Schweitzer,
%``Radial excitations of Q-balls, and their D-term,''
Phys. Rev. D \textbf{86} (2012), 096002
%%%%%%%%%%%%%%%%%%%%%%%%%%%%%%%%%%%%%%%%%%%%%%%%%%%%%%%%%%%%%%%%%%%%%%%%%%%%%%
%\cite{Loiko:2022noq}
\bibitem{Loiko:2022noq}
V.~Loiko and Y.~Shnir,
%``Q-ball stress stability criterion in U(1) gauged scalar theories,''
Phys. Rev. D \textbf{106} (2022) no.4, 045021
%%%%%%%%%%%%%%%%%%%%%%%%%%%%%%%%%%%%%%%%%%%%%%%%%%%%%%%%%%%%%%%%%%%%%%%%%%%%%%
%\cite{Panteleeva:2023aiz}
\bibitem{Panteleeva:2023aiz}
J.~Y.~Panteleeva,
%``Internal force distributions in the {\textquoteright}t Hooft-Polyakov monopole and Julia-Zee dyon,''
Phys. Rev. D \textbf{107} (2023) no.5, 055015
%%%%%%%%%%%%%%%%%%%%%%%%%%%%%%%%%%%%%%%%%%%%%%%%%%%%%%%%%%%%%%%%%%%%%%%%%%%%%%
%\cite{Farakos:2025byy}
\bibitem{Farakos:2025byy}
K.~Farakos, G.~Koutsoumbas, N.~E.~Mavromatos and A.~Zarafonitis,
%``On internal mechanical properties of Electroweak Magnetic Monopoles and their effects on stability,''
The European Physical Journal Special Topics (2025), 1-53.
%%%%%%%%%%%%%%%%%%%%%%%%%%%%%%%%%%%%%%%%%%%%%%%%%%%%%%%%%%%%%%%%%%%%%%%%%%%%%%
%\cite{Vakhitov:1973lcn}
\bibitem{Vakhitov:1973lcn}
N.~G.~Vakhitov and A.~A.~Kolokolov,
%``Stationary solutions of the wave equation in a medium with nonlinearity saturation,''
Radiophys. Quant. Electron. \textbf{16} (1973) no.7, 783-789
%%%%%%%%%%%%%%%%%%%%%%%%%%%%%%%%%%%%%%%%%%%%%%%%%%%%%%%%%%%%%%%%%%%%%%%%%%%%%%
%\cite{Kunz:2025tsy}
\bibitem{Kunz:2025tsy}
J.~Kunz, A.~Mikhaliuk and Y.~Shnir,
%``Two types of boson stars in the (3+1)-dimensional O(3) sigma-model,''
Phys. Rev. D \textbf{113} (2026) no.2, 025009
%%%%%%%%%%%%%%%%%%%%%%%%%%%%%%%%%%%%%%%%%%%%%%%%%%%%%%
%\cite{Leese:1989gi}
\bibitem{Leese:1989gi}
R.~A.~Leese, M.~Peyrard and W.~J.~Zakrzewski,
%``Soliton Scatterings in Some Relativistic Models in (2+1)-dimensions,''
Nonlinearity \textbf{3} (1990), 773-808
%%%%%%%%%%%%%%%%%%%%%%%%%%%%%%%%%%%%%%%%%%%%%%%%%%%%%%
%\cite{Salmi:2014hsa}
\bibitem{Salmi:2014hsa}
P.~Salmi and P.~Sutcliffe,
%``Aloof Baby Skyrmions,''
J. Phys. A \textbf{48} (2015) no.3, 035401
%%%%%%%%%%%%%%%%%%%%%%%%%%%%%%%%%%%%%%%%%%%%%%%%%%%%%%
%\cite{Gillard:2015eia}
\bibitem{Gillard:2015eia}
M.~Gillard, D.~Harland and M.~Speight,
%``Skyrmions with low binding energies,''
Nucl. Phys. B \textbf{895} (2015), 272-287
%%%%%%%%%%%%%%%%%%%%%%%%%%%%%%%%%%%%%%%%%%%%%%%%%%%%%%%%%%%%%%%%%%%%%%
%\cite{Mikhaliuk:2025mxy}
\bibitem{Mikhaliuk:2025mxy}
A.~Mikhaliuk and Y.~Shnir,
%``Boson stars in the $U(1)$ gauged $(3 + 1)$-dimensional $O(3)$ sigma model,''
Phys. Rev. D \textbf{112} (2025) no.2, 025015
%%%%%%%%%%%%%%%%%%%%%%%%%%%%%%%%%%%%%%%%%%%%%%%%%%%%%%%
%\cite{Loginov:2020xoj}
\bibitem{Loginov:2020xoj}
A.~Y.~Loginov and V.~V.~Gauzshtein,
%``Radially excited $U\left(1\right)$ gauged $Q$-balls,''
Phys. Rev. D \textbf{102} (2020) no.2, 025010
%%%%%%%%%%%%%%%%%%%%%%%%%%%%%%%%%%%%%%%%%%%%%%%%%%%%%%%%%%%%%%%%%%%%%%%%%%%%%%
%\cite{Laue:1911lrk}
\bibitem{Laue:1911lrk}
M.~Laue,
%``Zur Dynamik der Relativit{\"a}tstheorie,''
Annalen Phys. \textbf{340} (1911) no.8, 524-542
%%%%%%%%%%%%%%%%%%%%%%%%%%%%%%%%%%%%%%%%%%%%%%%%%%%%%%%%%%%%%%%%%%%%%%%%%%%%%%
%\cite{Pinto:2025plg}
\bibitem{Pinto:2025plg}
S.~R.~Pinto and P.~P.~Avelino,
%``Deviations from the von Laue condition: Implications for the on-shell Lagrangian of particles and fluids,''
Phys. Rev. D \textbf{111} (2025) no.8, 083556
%%%%%%%%%%%%%%%%%%%%%%%%%%%%%%%%%%%%%%%%%%
%%%%%%%%%%%%%%%%%%%%%%%%%%%%%%%%%%%%
%\cite{Perevalova:2016dln}
\bibitem{Perevalova:2016dln}
I.~A.~Perevalova, M.~V.~Polyakov and P.~Schweitzer,
%``On LHCb pentaquarks as a baryon-$\psi$(2S) bound state: prediction of isospin-$\frac3{2}$ pentaquarks with hidden charm,''
Phys. Rev. D \textbf{94} (2016) no.5, 054024
%%%%%%%%%%%%%%%%%%%%%%%%%%%%%%%%%%%%%%%%%%%%%%%%%%%%%%%%%%%%%%%%%%%%%%%%%%%%%%
%%%%%%%%%%%%%%%%%%%%%%%%%%%%%%%%%%%%%%%%%%%%%%%%%%%%%%%

%%%%%%%%%%%%%%%%%%%%%%%%%%%%%%%%%%%%%%%%%%%%%%%%%%%%%%%

\end{thebibliography}
\end{document}